\documentclass[aps,prd,amsmath,amsfonts,amssymb,eqsecnum,nofootinbib,
  floatfix,secnumarabic,preprintnumbers,superscriptaddress,nobalancelastpage,
  onecolumn,notitlepage,biblatex]{revtex4-1}
\usepackage{graphicx}
\usepackage{footnote,hyperref,xcolor}
\usepackage[bottom]{footmisc}
\usepackage{fancybox}
\usepackage{cancel}
\usepackage{diagbox}
\usepackage{caption}
\usepackage{subcaption}
\usepackage{multirow}         
\usepackage{blindtext}
\usepackage{enumitem}         
\usepackage[english]{babel}
\usepackage{natbib}
\usepackage{adjustbox}
\usepackage{cancel}
\def\be{\begin{equation}}
\def\ee{\end{equation}}

\def\beq{\begin{eqnarray}}
\def\eeq{\end{eqnarray}}

\begin{document}

\title{The 2HDM Doppelganger}

\author{Baradhwaj Coleppa}
 \email{baradhwaj@iitgn.ac.in}
\author{Agnivo Sarkar}%
 \email{agnivo.sarkar@iitgn.ac.in}
\affiliation{IIT Gandhinagar, Palaj Campus, Gujarat 382355, India}%

\date{\today}

\begin{abstract}
We discuss the structure of a model with an extended gauge symmetry group $SU(2)\times SU(2)\times U(1)$ with a correspondingly rich Electroweak Symmetry Breaking structure. In spite of the additional scalar degrees of freedom in the model, the presence of the extra gauge group $SU(2)$ and its associated heavy vector bosons ensures that the scalar spectrum of the model after symmetry breaking is identical to that of the Two Higgs Doublet Models. We construct the model and discuss its implications, specifically the phenomenology associated with this class of models in contrast to the 2HDM.
\end{abstract}

\pacs{Valid PACS appear here}
\maketitle

\section{Introduction}
While the discovery of a Higgs boson whose properties are largely consistent with that predicted in the Standard Model (SM) \cite{Aad:2012tfa,Chatrchyan:2012ufa,Khachatryan:2016vau,ATLAS:2019slw,Sirunyan:2018sgc,Sirunyan:2018koj} is undoubtedly a significant achievement of the ATLAS and CMS experiments of the Large Hadron Collider (LHC), vexing questions surrounding the stability of the Higgs boson mass, explanation of neutrino mass, and the origin of dark matter still remain. Generic ``Beyond the Standard Model" (BSM) scenarios that have been constructed to answer one or more of such outstanding theoretical questions usually necessitate the introduction of additional degrees of freedom that should potentially be discovered at the LHC. BSM scenarios typically involve enlarging the gauge structure of the SM, invoking non-trivial patterns of Electroweak Symmetry Breaking (EWSB) involving additional scalar particles, or enlarging the matter content of the SM - many BSM scenarios admit more than one of these possibilities. The success or failure of these models would ultimately be decided by the confirmation of their predictions at the LHC. Given that there are a large number of such BSM scenarios available, typically phenomenologists and experimentalists tend to concentrate of explaining the results of the ATLAS and CMS data (null results or otherwise) in the context of certain simplified scenarios that have features that are rather generic. One such example in the context of extended models with extended scalar sectors is the Two Higgs Doublet Model (2HDM) \cite{Lee:1973iz,Akeroyd:1998dt,Aoki:2009ha,Baak:2011ze,Posch:2010hx,Kominis:1994fa} (also see \cite{Branco:2011iw} for a comprehensive review).

The 2HDM extends the SM by invoking a second Higgs doublet, and EWSB is engineered by both the doublets via their vacuum expectation values (vev) $v_1$ and $v_2$. After symmetry breaking, one spin-0 linear combination of the fields is identified with the SM Higgs boson with mass 125 GeV, and the second Higgs with different couplings to the gauge bosons and fermions, can be either heavier or lighter than the SM one. There is a huge experimental interest in discovering these extra Higgs bosons at the ATLAS \cite{Aad:2019zwb} and CMS \cite{Sirunyan:2019wph,Sirunyan:2019pqw} experiments, partly because the scalar sectors of the (Type II) 2HDM is identical to that of the Minimal Supersymmetric Standard Model (MSSM). While the search for such extra Higgs bosons is generic in the sense that experiments typically look for a particular final state, the interpretation of the findings of these experiments in a model-dependent scenario can have wildly different results depending on the couplings of these extra Higgs bosons, and their production cross-section and decay branching ratios. In this context, in this paper we present the details of a model that has two interesting features: \emph{i)} an extended electroweak gauge group $SU(2)\times SU(2)\times U(1)$, and \emph{ii)} a scalar spectrum that is identical to that of the 2HDM. The model is a variant of that presented in Ref.~\cite{Chivukula:2009ck,Chivukula:2011ag} as the ``Top Triangle Moose Model", with one of the non-linear sigma model fields being replaced by a linear Higgs field. As we demonstrate below, this changes the scalar structure of the model in such a way as to mimic the 2HDM (and MSSM) scenarios, and if the associated gauge bosons are heavy, this model would look indistinguishable from the 2HDM at low energies. The goal of the present work is to demonstrate the features of this model, and sketch the rich and varied phenomenology this alternate scheme of EWSB allows us to pursue.

This paper is organized as follows: in Sec.~\ref{sec:model}, we present the details of the model mostly restricting to those aspects of it that are distinct from that in Refs.~\cite{Chivukula:2009ck,Chivukula:2011ag}. In Sec.~\ref{sec:scalar_couplings}, we compute the couplings of the neutral and charged Higgs bosons in the theory and identify possibilities for distinct LHC signatures. In Sec.~\ref{sec:pheno}, we analyze the parameter space of the model that survives the LHC experimental constraints, and present the cross-sections and decay branching ratios of the Higgs bosons before concluding in Sec.~\ref{sec:conculsions}.
\section{The Model}
\label{sec:model}
\begin{center}
\begin{figure}
\includegraphics[scale=1.0]{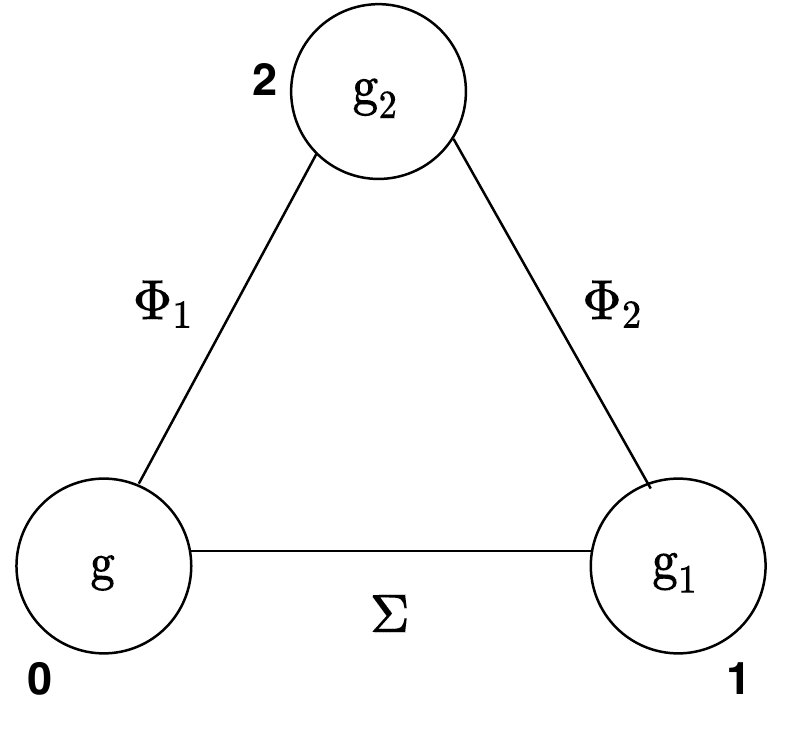}
\caption{The model under consideration in the ``moose" notation. ``Sites" 0 and 2 contain the (mostly) SM gauge group $SU(2)\times U(1)$ with the other $SU(2)$ residing at site 2. The gauge couplings of the $SU(2)\times SU(2)\times U(1)$ groups are denoted by $g_0,g_1$ and $g_2$ respectively. Symmetry breaking in this model is engineered by two Higgs doublets $\Phi_{1,2}$ and a non-linear sigma model field $\Sigma$.}
\label{fig:moose}
\end{figure}
\end{center}

The electroweak gauge structure of the model is a minimal extension of the SM with the ``221" structure of gauge groups \cite{Chivukula:2006cg,Abe:2012fb,Coleppa:2018fau} $SU(2)_0\times SU(2)_1\times U(1)$. Spontaneous symmetry breaking in the model is engineered via two Higgs doublets $\Phi_1$ and $\Phi_2$ and a non-linear sigma model field $\Sigma$. We illustrate this in Fig.~\ref{fig:moose} using the moose notation with the two $SU(2)$ groups residing at ``sites" 0 and 1 and the $U(1)$ group at site 2. 
The non-linear sigma field $\Sigma$ breaks the $SU(2)_0\times SU(2)_1$ down to the diagonall $SU(2)_V$. The two additional scalar doublets in the model transform as ($\textbf{2},\textbf{1},\frac{\textbf{1}}{\textbf{2}}$) and ($\textbf{1},\textbf{2},\frac{\textbf{1}}{\textbf{2}}$) respectively under the full symmetry group of the model. We denote the vacuum expectation values (vevs) of the different scalar fields as $\langle \Phi_{1} \rangle = \frac{f_{1}}{\sqrt{2}}$,  $\langle \Phi_{2} \rangle = \frac{f_{2}}{\sqrt{2}}$ and $\langle \Sigma \rangle = F$ and follow the same parametrization as that in Refs.~\cite{Chivukula:2009ck,Chivukula:2011ag}:\footnote{While Refs.~\cite{Chivukula:2009ck,Chivukula:2011ag} treated $\sin\beta$ as a small parameter with the EWSB mostly engineered via a ``higgsless" mechanism, in the present paper we place no such constraints on $\sin\beta$ at the outset.}
\be f_{1} = v\sin\beta; \,\,F = f_{2} = \sqrt{2}v\cos\beta, \label{eqn:decayF}\ee
where $v$ is the vev of the Higgs field in the SM (246 GeV). The explicit form of the fields are given as
\be
\Sigma=  \exp{\left(\frac{i\Pi^{a}_{\Sigma}\sigma^{a}}{F}\right)}, \,\, \Phi_{i} = 
\begin{pmatrix}
(f_{i} + H_{i} + i\Pi^{0}_{i})/\sqrt{2} \\
i\Pi^{-}_{i}
\end{pmatrix};\,
i = 1, 2,
\ee
where $\sigma^{a}$ are the three Pauli matrices. The gauge couplings for the $SU(2)_0$, $SU(2)_1$, and $U(1)$ groups are parametrized as
\be 
g=\frac{e}{\sin\theta\cos\phi};\,\, g_{1}=\frac{e}{\sin\theta\sin\phi},\,\,g_{2}=\frac{e}{\cos\theta}.
\label{eq:parametrize}
\ee
\subsection{Symmetry Breaking}

To begin discussing the symmetry breaking structure of this particular 221 model, we write down the most general potential for the scalar fields consistent with the gauge symmetries:
\begin{align}
\begin{split}
V(\Phi_{1},\Phi_{2},\Sigma) & = m^{2}_{11}\Phi^{\dagger}_{1}\Phi_{1} + m^{2}_{22}\Phi^{\dagger}_{2}\Phi_{2} - (m^{2}_{12}\Phi^{\dagger}_{1}\Sigma\Phi_{2} + \textnormal{h.c.}) + \frac{\beta_{1}}{2}(\Phi^{\dagger}_{1}\Phi_{1})^{2} + \frac{\beta_{2}}{2}(\Phi^{\dagger}_{2}\Phi_{2})^{2} \\ 
& + \beta_{3}(\Phi^{\dagger}_{1}\Phi_{1})(\Phi^{\dagger}_{2}\Phi_{2}) + \beta_{4}(\Phi^{\dagger}_{1}\Sigma\Phi_{2})(\Phi^{\dagger}_{2}\Sigma^{\dagger}\Phi_{1}) + \bigg[\frac{\beta_{5}}{2}(\Phi^{\dagger}_{1}\Sigma\Phi_{2})^{2} + \textnormal{h.c.}\bigg].
\end{split}
\end{align}
Through the following identification of parameters
\begin{align}
\begin{split}
m^{2}_{11}& = - \bigg[\lambda_{1}f^{2}_{1} + \lambda_{3}(f^{2}_{1} + f^{2}_{2})\bigg]; \,m^{2}_{22} = - \bigg[ \lambda_{2}f^{2}_{2} + \lambda_{3}(f^{2}_{1} + f^{2}_{2})\bigg];\, m^{2}_{12} = \lambda_{5}\frac{f_{1}f_{2}}{2}, \beta_{1} = 2(\lambda_{1} + \lambda_{3});\\
\beta_{2}& = 2(\lambda_{2} + \lambda_{3});\,\beta_{3} = (\lambda_{4} + 2\lambda_{3});\,\beta_{4} = \frac{1}{2}(\lambda_{5} + \lambda_{6}) - \lambda_{4};\,\beta_{5} = \frac{1}{2}(\lambda_{5} - \lambda_{6}),
\end{split}
\label{eqn:L1}
\end{align}
we could recast the Lagrangian to a form more commonly used in the 2HDM literature\footnote{Note, however, that the $\Phi_{1,2}$ are charged under different gauge groups unlike in the 2HDM; the combination $\Phi_1^{\dagger}\Phi_2$ is \emph{not} gauge invariant in the present model while the term $\Phi_1^{\dagger}\Sigma\Phi_2$ is.}:
\begin{align}
\begin{split}
V(\Phi_{1},\Phi_{2},\Sigma) & =  \lambda_{1}\bigg[\Phi^{\dagger}_{1}\Phi_{1} - \frac{f^{2}}{2}\bigg]^{2} + \lambda_{2}\bigg[\Phi^{\dagger}_{2}\Phi_{2} - \frac{F_{2}}{2}\bigg]^{2} + \lambda_{3}\bigg[\Phi^{\dagger}_{1}\Phi_{1} + \Phi^{\dagger}_{2}\Phi_{2} - \frac{f^{2} + F^{2}}{2}\bigg]^{2} \\
& + \lambda_{4}\bigg[(\Phi^{\dagger}_{1}\Phi_{1})(\Phi^{\dagger}_{2}\Phi_{2}) - (\Phi^{\dagger}_{1}\Sigma\Phi_{2})(\Phi^{\dagger}_{2}\Sigma^{\dagger}\Phi_{1})\bigg] + \lambda_{5}\bigg[\textnormal{Re}(\Phi^{\dagger}_{1}\Sigma\Phi_{2}) - \frac{f F}{2}\bigg]^{2} \\
& + \lambda_{6} \,\textnormal{Im}\bigg[\Phi^{\dagger}_{1}\Sigma\Phi_{2}\bigg]^{2},
\end{split}
\label{eqn:L2}
 \end{align}
where use has been made use of Eqn.~\ref{eqn:decayF}, and we have relabeled $f_1\to f$ for notational simplicity. We will fix all $\lambda_{i}$'s to be real parameters to insure the hermiticity of the Lagrangian. Out of the eleven scalar degrees of freedom, six are Goldstone modes that get eaten up by the massive gauge bosons while five remain as physical scalar particles in the spectrum. In keeping with the standard literature, these will be denoted as the charged Higgs bosons $H^{\pm}$, two CP-even Higgses  $(H, h)$ and a pseudo-scalar $A$. The mass matrix of the two CP-even states can be written as
 \begin{equation}
M^{2}_{h,H} = 
\begin{bmatrix}
2(\lambda_{1} + \lambda_{3})f^{2} + \frac{\lambda_{5}}{2}F^{2}  & 2(\lambda_{3} + \frac{\lambda_{5}}{4})fF \\
      & \\
2(\lambda_{3} + \frac{\lambda_{5}}{4})fF & 2(\lambda_{2} + \lambda_{3})F^{2} + \frac{\lambda_{5}}{2}f^{2} 
\end{bmatrix}
\end{equation}
with eigenvalues
\begin{align}
\begin{split}
m^{2}_{H} &= \frac{1}{2}\bigg[(A_{s} + C_{s}) + \sqrt{(A_{s} - C_{s})^{2} + 4B^{2}_{s}}\bigg]\\
m^{2}_{h} &= \frac{1}{2}\bigg[(A_{s} + C_{s}) - \sqrt{(A_{s} - C_{s})^{2} +4B^{2}_{s}}\bigg] ,
\end{split}
\end{align}
where \[ \qquad A_{s} = 2(\lambda_{1} + \lambda_{3})f^{2} + \frac{\lambda_{5}}{2}F^{2},\,B_{s} = 2(\lambda_{3} + \frac{\lambda_{5}}{4})fF,\, \textrm{and}\, \,C_{s} = 2(\lambda_{2} + \lambda_{3})F^{2} + \frac{\lambda_{5}}{2}f^{2}. \]
The physical CP-even Higgs eigenstates are 
\begin{align}
\begin{split}
H &= \cos\alpha\, H_{1} + \sin\alpha\, H_{2} \\
h &= - \sin\alpha\, H_{1} + \cos\alpha\, H_{2},
\end{split}
\end{align}
with the mixing angle $\alpha$ defined to be 
\be\sin2\alpha = \frac{2B_{s}}{\sqrt{(A_{s} - C_{s})^{2} + 4B^{2}_{s}}}. \ee
We will identify the $h$ with the SM-like Higgs of mass 126 GeV, while the $H$ would be a heavier Higgs. We will detail the production and decay of these states in Sec.~\ref{sec:pheno}.
Notice that while the mass term for the CP-even Higgses can arise from terms like $\Phi_i^{\dagger}\Phi_i$, the $H^{\pm}$ mass comes solely from  the $\lambda_{4}$ term in Eqn.~\ref{eqn:L2} - the $\lambda_{5}$ and $\lambda_{6}$ terms do not contain any terms quadratic in the pions and describe interactions. The mass matrix for the charged scalars thus takes the following form:
\begin{equation}
M^{2}_{\pi^{\pm}} = \frac{\lambda_{4}}{2}
\begin{bmatrix}
f^{2} & - fF & f^2 \\
-fF& F^{2} & - fF \\
f^2 & - fF & f^{2}
\end{bmatrix}.
\end{equation}
The corresponding eigenstates of the Goldstone modes $G^{\pm}_{1}$, $G^{\pm}_{2}$ which get eaten by the $W^{\pm}_{\mu}$ and $W^{'\pm}_{\mu}$ are given by
\begin{align}
\begin{split}
G_{1}^{\pm}  &= - \frac{1}{\sqrt{2}}\Pi_{\Sigma}^{\pm}  + \frac{1}{\sqrt{2}}\Pi_{2}^{\pm} \\
G_{2}^{\pm}  &= \frac{F}{2v}\Pi_{\Sigma}^{\pm}  +  \frac{f}{v}\Pi_{1}^{\pm} + \frac{F}{2v}\Pi_{2}^{\pm} ,
\label{eqn:goldstone}
\end{split}
\end{align}
while the physical charged Higgs corresponds to the combination 
\be 
H^{\pm}   = \frac{f}{\sqrt{2}v}\Pi_{\Sigma}^{\pm}   -  \frac{F}{\sqrt{2}v}\Pi_{1}^{\pm}  + \frac{f}{\sqrt{2}v}\Pi_{2}^{\pm} 
\label{eqn:chargedHiggs}
\ee
 with the corresponding mass eigen-value is $M^{2}_{H^{\pm}}  = \frac{\lambda_{4}}{2}(2f^{2} + F^{2})$. 
The neutral Goldstone mode $G^0$ and the physical pseudoscalar $A$ can be obtained by the following replacements in Eqns.~\ref{eqn:goldstone}, \ref{eqn:chargedHiggs}: 
\[ \Pi^{\pm}_{\Sigma} \rightarrow \Pi^{0}_{\Sigma},\, \,\Pi^{\pm}_{1} \rightarrow \Pi^{0}_{1},\,\, \textrm{and}\, \, \,\Pi^{\pm}_{2} \rightarrow \Pi^{0}_{2}. \]

We note that the scalar sector of our model as parametrized in Eqn.~\ref{eqn:L2} requires six free parameters (the $\lambda_{i}$) in addition to the two vevs, so we need a total of eight parameters to write down the most general potential invariant under $SU(2)_0\times SU(2)_1 \times U(1)$. We can trade two of the eight parameters to $\beta$ which is the ratio of two vev's and the mixing angle $\alpha$ between the two CP-even Higgs eigenstates. Five of the other six  $\lambda_{i}$s can be expressed in terms of the five physical Higgs masses and the EWSB scale $v$. These expressions come in handy when imposing theoretical and experimental constraints on the scalar couplings and we present them below.

\begin{subequations}
\begin{eqnarray}
\lambda_{1} &=& \frac{1}{2v^{2}\sin^{2}\beta}\bigg[(m^{2}_{H}\cos^{2}\alpha + m^{2}_{h}\sin^{2}\alpha) - \frac{1}{\sqrt{2}}\tan\beta\sin\alpha\cos\alpha(m^{2}_{H} - m^{2}_{h})\bigg] - \frac{\lambda_{5}}{2}(2\cot^{2}\beta - 1). \\
\lambda_{2} &=& \frac{1}{4v^{2}\sin^{2}\beta}\bigg[(m^{2}_{H}\cos^{2}\alpha + m^{2}_{h}\sin^{2}\alpha) - \frac{\sqrt{2}}{\tan\beta}\sin\alpha\cos\alpha(m^{2}_{H} - m^{2}_{h})\bigg] - \frac{\lambda_{5}}{4}\bigg(\frac{\tan^{2}\beta}{2} - 1\bigg). \\
\lambda_{3} &=& \frac{\sin\alpha\cos\alpha}{2\sqrt{2}v^{2}\sin\beta\cos\beta}(m^{2}_{H} - m^{2}_{h}) - \frac{\lambda_{5}}{2}. \\
\lambda_{4} &=& \frac{m^{2}_{H^{\pm}}}{v^{2}}. \\
\lambda_{6} &=& \frac{m^{2}_{A}}{v^{2}}.
\end{eqnarray}
\end{subequations}

\subsection{Gauge Sector}
The kinetic energy part of our $SU(2)\times SU(2)\times U(1)$ model can be written down in the usual canonically normalized form
\begin{equation}
\mathcal{L}_{K.E} = -\frac{1}{4}\sum_{i=0}^{1}F^{a}_{i\mu\nu}F^{a\mu\nu}_{i}  - \frac{1}{4}B_{\mu\nu}B^{\mu\nu},
\end{equation}
where $F^{a\mu\nu}_{i}$ is the energy-momentum tensor for the non-abelian gauge fields associated with the two $SU(2)$ gauge groups and $B_{\mu\nu} = \partial_{\mu}B_{\nu} - \partial_{\nu}B_{\mu}$ is the Abelian $U(1)$ counterpart. After spontaneous symmetry breaking six out of seven gauge bosons become massive. To understand the gauge boson masses, we begin by writing down the gauge invariant kinetic energy terms of the scalar fields:
\begin{equation}
\mathcal{L} = \frac{F^{2}}{2}\textnormal{Tr}[D_{\mu}\Sigma^{\dagger}D^{\mu}\Sigma] + D_{\mu}\Phi^{\dagger}_{1}D^{\mu}\Phi_{1} + D_{\mu}\Phi^{\dagger}_{2}D^{\mu}\Phi_{2}.
\label{eq:gauge}
\end{equation}
The covariant derivatives are given by
\begin{align}
\begin{split}
D_{\mu}\Sigma & =  \partial_{\mu}\Sigma + ig\tilde{W}_{0\mu}\Sigma - ig_{1}\Sigma \tilde{W}_{1\mu}  \\ 
 D_{\mu}\Phi_{1}& =  \partial_{\mu}\Phi_{1} + ig\tilde{W}_{0\mu}\Phi_{1} -\frac{ig_{2}}{2}B_{2\mu}\Phi_{1} \\
 D_{\mu}\Phi_{2} & = \partial_{\mu}\Phi_{2} + ig_{1}\tilde{W}_{1\mu}\Phi_{2} - \frac{ig_{2}}{2}B_{2\mu}\Phi_{2},
\label{eq:covariants}
\end{split}
\end{align}
where the matrix fields $\tilde{W}_{0\mu} = \frac{W^{a}_{0\mu}\sigma^{a}}{2}$ and $\tilde{W}_{1\mu} = \frac{W^{a}_{1\mu}\sigma^{a}}{2}$. Following Refs.~\cite{Chivukula:2009ck,Chivukula:2011ag}, we will treat $\sin\phi$ in Eqn.~\ref{eq:parametrize} as a small parameter and diagonalize the gauge boson mass matrices perturbatively in $\sin\phi$ (which we will relabel as $x$ henceforth for notational simplicity). We will not reproduce the calculations already laid out in Refs.~\cite{Chivukula:2009ck,Chivukula:2011ag} here - the salient features are the existence of heavy $W'$ and $Z'$ bosons whose wavefunctions are peaked away from the $SU(2)_0$ and the $U(1)$ sites and the SM-like $W$ and $Z$ bosons that do not have a significant overlap with the $SU(2)_1$ gauge group. For instance, the wave functions of the charged gauge bosons are
\begin{align}
\begin{split}
W^{\pm}_{\mu} &= (1 - \frac{x^{2}}{8})W^{\pm}_{0\mu} + \frac{x}{2}W_{1\mu}\\
W^{'\pm}_{\mu} &= - \frac{x}{2}W_{0\mu} + (1 - \frac{x^{2}}{8})W_{1\mu}
\end{split}
\label{eq:Wwavefn}
\end{align}
The photon remains massless and is given by
\begin{equation}
A_{\mu} = \frac{e}{g} W^{3}_{0\mu} + \frac{e}{g_{1}}W^{3}_{1\mu} + \frac{e}{g_{2}}B_{2\mu}
\label{eq:photon}
\end{equation}
with the different coupling constants related by
\begin{equation}
\frac{1}{e^{2}} = \frac{1}{g^{2}} + \frac{1}{g^{2}_{1}} + \frac{1}{g^{2}_{2}}.
\end{equation}
One could proceed to evaluate the various self-couplings between the gauge fields - we relegate the results to Appendix \ref{app: gauge}.
\subsection{Fermion Sector}
The model admits the SM-like fermions and also heavy vector-like fermions. Charge assignments of the various species under the different gauge groups can, in principle, be done in a number of ways. The assignment that we will adopt is motivated by two factors:
\begin{enumerate}[label=(\roman*)]
\item That the overall structure of the model stays as close to possible to the original version presented in Refs.~\cite{Chivukula:2009ck,Chivukula:2011ag}.
\item That the spectrum and the pattern of couplings mimic one of the 2HDMs to the extent possible so that a meaningful comparison can be made.
\end{enumerate}
Admittedly one or both of these constraints can be relaxed that will lead to different possibilities, but in this paper we will operate within the confines of the aforementioned points. Notice that even $(ii)$ offers multiple choices with regard to the couplings of the fermions to the scalar fields. We will choose to follow the conventions of the so-called Type I 2HDM wherein both up-type and down-type quarks and the charged leptons couple to one of the two Higgs doublets, though both of them take part in EWSB. However, we mention at the outset that any comparison to 2HDM cannot be one-to-one as the particle content of this model in the gauge and fermionic sectors is richer than that of the 2HDM.

The left-handed fermions are $SU(2)$ doublets which reside at site 0 and 1 - we will denotes these as $\psi_{L0}$ and $\psi_{L1}$. In addition to the SM-like $u_R$ and $d_R$ at site 2, there are also right-handed fermions $\psi_{R2}$. that are doublets under $SU(2)_1$. The $\psi_{L0}$, $\psi_{L1}$, and $\psi_{R1}$ have $U(1)$ charges similar to the SM doublets: $\frac{1}{6}$ for quarks and $-\frac{1}{2}$ for leptons. The right-handed up quark has a $U(1)$ charge $\frac{2}{3}$ while the down type quark carries $- \frac{1}{3}$. For the case of leptons, the hypercharge assignments are similar to the SM. We summarize all the electroweak charges for the fermions in the Table \ref{tab:charges}. 
\begin{table}[h!]
\centering
\begin{tabular}{| c | c | c | c |}
\hline 
    & $\psi_{L0}$ & $\psi_{L1}$ , $\psi_{R1}$ & $u_{R2}, d_{R2}$ \\
 \hline  \hline 
 $SU(2)_0$ & 2 & 1 & 1 \\
 \hline 
 $SU(2)_1$ & 1 & 2 & 1 \\
 \hline 
$U(1)$ & $\frac{1}{6}$ & $\frac{1}{6}$ & $\frac{2}{3}$ or $- \frac{1}{3}$ \\
 \hline   
\end{tabular}
\caption{Fermionic charge assignments in the model - in addition to the SM-like left-handed doublets and right-handed $SU(2)$ singlets, there is also a right-handed fermion doublet under $SU(2)_1$.}
 \label{tab:charges}
\end{table} 

The SM fermions (predominantly) derive their masses from coupling to the $\Phi_1$ field. In view of this, we can write down the following term in the fermion sector of the Lagrangian:
\be
\mathcal{L} = \lambda^{ij}_{u0}\bar{\psi}_{L0}\Phi_{1}u_{R2} +\lambda^{ij}_{d0}\bar{\psi}_{L0}\tilde{\Phi}_{1}d_{R2}+ h.c.,
\ee
where $\lambda^{ij}_{u0},\lambda^{ij}_{d0}$ are the Yukawa couplings that are set by the masses of the up and the down-type quarks respectively and $\tilde{\Phi}_{1}=-i\sigma_2\Phi_1$. The vector-like fermions at site 1 are unaffected by EWSB and thus admit a Dirac mass term $\bar{\psi}_{L1}\psi_{R1}$. In addition, there are two gauge invariant dimension-4 terms that we cannot ignore: the coupling of the $\psi_{L0}$ to the $u_R,d_R$ via the link field $\Sigma$ (which was the primary mass term for the fermions in the Higgsless mechanism in Refs.~\cite{Chivukula:2009ck,Chivukula:2011ag}) and  $\bar{\psi}_{L1}\Phi_{2}f_{R2}$, where $f_{R2}$ is either an up or a down type fermion residing at site 2. The latter is a new term quite distinct from the Type I 2HDM that can arise here due to the presence of the extra gauge group and the presence of vector-like fermions. Putting everything together, the mass terms for the various quarks and leptons in the model are given by the following terms in the Lagrangian: 
\be
\mathcal{L} = - \lambda^{ij}_{u0}\bar{\psi}_{L0}\Phi_{1}u_{R2} -\lambda^{ij}_{d0}\bar{\psi}_{L0}\tilde{\Phi}_{1}d_{R2} - \lambda_{L}\bar{\psi}_{L0}\Sigma \psi_{R1} - M_{D}\bar{\psi}_{L1}\psi_{R1} - \lambda_{uR}\bar{\psi}_{L1}\Phi_{2} u_{R2}-\lambda_{dR}\bar{\psi}_{L1}\tilde{\Phi}_{2} d_{R2} + \textrm{h.c.},
\label{eq:fermionL}
\ee
where we have written down the coupling of $\Phi_2$ with the up and down type fermions with strengths $\lambda_{uR},\lambda_{dR}$. In order to mimic the Type I 2HDM to the extent possible, the masses of all SM fermions arise from the $\lambda^{ij}_{u0},\lambda^{ij}_{d0}$ terms and thus we will assume this subdominant contribution does not affect the up-down mass splitting appreciably. \footnote{We use $u$ and $d$ generically to mean any up-type or down-type quark.}. Thus at the outset, we assume the hierarchy of coupling strengths $\lambda_{uR,dR}\ll \lambda_{L}<\lambda_{u0,d0}$\footnote{While this is certainly not necessary \emph{a priori}, this choice will enable us to write down the condition for ideal fermion delocalization in this model identical to Ref.~\cite{Chivukula:2009ck} - see Eqn.~\ref{eq:IDF}.}. This ensures that any deviations in the Higgs-fermion interaction strengths compared to the SM arise predominantly from the non-linear sigma model fields. To satisfy FCNC constraints we have chosen $\lambda^{ij}_{L,R} = \lambda_{L,R}\delta^{ij}$ and $M^{ij}_{D} = M_{D}\delta^{ij}$ and thus all the non-trivial flavor structure of the model is encoded in $\lambda^{ij}_{u0},\lambda^{ij}_{d0}$. To quantify the mixing between the SM fermions and their heavier partners, we define the following parameters:
\be
\epsilon_{L} = \frac{\lambda_{L}}{M_{D}};\, \epsilon_{fR} = \frac{\lambda_{fR}F}{M_{D}},
\ee
where $f$ is either $u$ or $d$. For notational simplicity, we define
\be
a_f = \frac{\lambda_{f0}v\sin\beta}{\sqrt{2}M_{D}}.
\ee
The fermionic mass matrix takes the form 
\begin{equation}
M_{F} = M_{D}\begin{bmatrix}
\epsilon_{L} & a_f \\
1 & \epsilon_{R}
\end{bmatrix}
\end{equation} 
We require that the dominant contribution to the SM fermion mass arises out of its coupling to the Higgs field $\Phi_1$. Accordingly, diagonalizing the matrix pertubatively in $\epsilon_{L}$ and $\epsilon_{R}$, we get the light fermion eigenvalue
\begin{equation} 
m_{f} = M_{D}a_f\bigg[ 1 + \frac{\epsilon^{2}_{L} + \epsilon^{2}_{R} + \frac{2}{a_{f}}\epsilon_{L}\epsilon_{R}}{2(-1 + a_f^{2})} \bigg].
\label{eq:fermionmass}
\end{equation} 
In the limit $\epsilon_{L,R}\to 0$, choosing the Yukawa coupling $a_f$ appropriately yields the masses of the light fermions. For completeness, we write down the mass the heavy partner below:
\begin{equation}
m_{F} = M_{D}\bigg[1 - \frac{\epsilon^{2}_{fL} + \epsilon^{2}_{fR} + 2a_f\epsilon_{L}\epsilon_{R}}{2(-1 + a_f^{2})} \bigg]
\end{equation}
The wave functions of the left and right handed SM fermion are:
\begin{align}
\begin{split}
f_{L} &= f^{0}_{L}\psi^{f}_{L0} + f^{1}_{L}\psi^{f}_{L1} \\
&= \bigg(1 - \frac{(\epsilon_{L} + a_f\epsilon_{R})^{2}}{2(-1 + a_f^{2})^{2}}\bigg)\psi_{L0} + \bigg(\frac{\epsilon_{L} + a_f\epsilon_{R}}{-1 + a_f^{2}}\bigg)\psi_{L1} \\
f_{R} &= f^{1}_{R}\psi^{f}_{R1} + f^{2}_{R}\psi^{f}_{R2} \\
&= \bigg(\frac{a_f\epsilon_{L} + \epsilon_{R}}{-1 + a_f^{2}}\bigg)\psi_{R1} + \bigg(1 - \frac{(a_f\epsilon_{L} + \epsilon_{R})^{2}}{2(-1 + a_f^{2})^{2}} \bigg)\psi_{R2}
\end{split}
\label{eq:fermionwavefn}
\end{align}
and those of the heavy partners are orthogonal combinations of the above.

With all the mass eigenstates in place, we can compute the couplings of the SM and heavy fermions with the gauge sector. For instance, the coupling of  $W^{\pm}$ with $tb$ is calculated as    
\be
g^{L}_{Wtb} = gv^{0}_{W}t^{0}_{L}b^{0}_{L} + g_{1}v^{1}_{W}t^{1}_{L}b^{1}_{L},
\ee
where $v_{W}^i,t_L^i,b_L^i$ are the amount of of the $W,t_L,b_L$ wavefunctions at site $i$. Plugging in these values from Eqns.~\ref{eq:Wwavefn} and \ref{eq:fermionwavefn} and taking $\sin\theta =\sin\theta_{W}\big(1 + \frac{x^{2}}{8}\big)$ \footnote{This is evaluated using the definition of the Weinberg angle $\cos^2\theta_w=\frac{M_W^2}{M_Z^2}$.}, we get
\be
g^{Wud}_{L} = \frac{e}{\sin\theta_{w}}\bigg[1 + \frac{x^{2}}{4} - \frac{1}{2}\frac{(\epsilon_{L} + a_f\epsilon_{R})^{2}}{(-1 + a_f^{2})^{2}} \bigg].
\ee
Following the argument below Eqn.~\ref{eq:fermionL}, we neglect the $\epsilon_{R}$ contribution \footnote{Note that the contribution to the $\rho$ parameter from the heavy top-bottom is exactly as computed in Ref.~\cite{Chivukula:2006cg} and reads $\Delta\rho = \frac{M^{2}_{D}\epsilon^{4}_{fR}}{16\pi^{2}v^{2}}$. However, since the $\epsilon_{fR}$ is not the dominant contribution to the light fermion mass, this can be tuned as small as necessary to evade the constraints.}and rewrite the coupling as
\be
g^{Wud}_{L} = \frac{e}{\sin\theta_{w}}\bigg[1 + \frac{x^{2}}{4} - \frac{\epsilon^{2}_{L}}{2} \bigg].
\ee
Thus the condition for the ideal fermion delocalization which renders the $Wud$ coupling SM-like while concurrently making the $W'ud$ coupling zero can be written in exactly the same fashion as in Refs.~\cite{Chivukula:2009ck,Chivukula:2011ag}:\footnote{We will also assume that the top quark is also declocalized in exactly the same fashion because of constraints arising from $Z\bar{b}_Lb_L$ coupling - see Ref.~\cite{Chivukula:2009ck} for details.}
\begin{equation}
\epsilon^{2}_{L} = \frac{x^{2}}{2}.
\label{eq:IDF}
\end{equation}
We summarize the gauge-fermion couplings in Appendix \ref{app: gauge}.
\section{Scalar Sector}
\label{sec:scalar_couplings}
In this section, we will present the couplings of the charged and neutral Higgs bosons in the model to the gauge and fermion sectors. We will then briefly discuss some of the constraints arising due to the demand of vacuum stability on the Lagrangian parameters. 

\subsection{Couplings}
We begin by computing the couplings of the Higgs bosons in the model to the gauge bosons - there are two basic kinds of three point vertices: $VVX$ and $VXX$, where $V$ generically means a (light or heavy) gauge boson and $X$ is a (neutral or charged) higgs boson. These couplings arise from the kinetic energy terms in the Lagrangian \ref{eq:gauge} as usual. We begin by tabulating all couplings of the first type in Table \ref{tab:VVX}.

\begin{table}[h!]
\begin{center}
\resizebox{12cm}{!}
{
\begin{tabular}{|   c   ||   c   |}
\hline
Vertex & Strength \\
\hline \hline
$hW^{+}_{\mu}W^{-}_{\nu}$ & $\frac{e^{2}v}{8\sin^{2}\theta_{w}}\bigg(\big(\sqrt{2}\cos\alpha\cos\beta - 4\sin\alpha\sin\beta\big) - \frac{x^{2}}{4}\big(\sqrt{2}\cos\alpha\cos\beta + 8\sin\alpha\sin\beta\big)\bigg)$ \\
\hline
$HW^{+}_{\mu}W^{-}_{\nu}$ & $\frac{e^{2}v}{8\sin^{2}\theta_{w}}\bigg(\big(\sqrt{2}\cos\beta\sin\alpha + 4\cos\alpha\sin\beta\big) - \frac{x^{2}}{4}\big(\sqrt{2}\cos\beta\sin\alpha - 8\cos\alpha\sin\beta\big)\bigg)$ \\
\hline
$hW^{'+}_{\mu}W^{'-}_{\nu}$ & $\frac{e^{2}v\cos\alpha\cos\beta}{\sqrt{2}x^{2}\sin^{2}\theta_{w}}\bigg(1 - \frac{x^{2}}{2}\bigg)$ \\
\hline
$HW^{'+}_{\mu}W^{'-}_{\nu}$ & $\frac{e^{2}v\cos\beta\sin\alpha}{\sqrt{2}x^{2}\sin^{2}\theta_{w}}\bigg(1 - \frac{x^{2}}{2}\bigg)$ \\
\hline
$hW^{+}_{\mu}W^{'-}_{\nu}$ & $\frac{e^{2}v}{2\sqrt{2}x\sin^{2}\theta_{w}}\bigg(\cos\alpha\cos\beta + \frac{x^{2}}{2}\big(\sqrt{2}\sin\alpha\sin\beta - \frac{3}{4}\cos\alpha\cos\beta\big)\bigg)$ \\
\hline
$HW^{+}_{\mu}W^{'-}_{\nu}$ & $\frac{e^{2}v}{2\sqrt{2}x\sin^{2}\theta_{w}}\bigg(\sin\alpha\cos\beta - \frac{x^{2}}{2}\big(\sqrt{2}\cos\alpha\sin\beta + \frac{3}{4}\cos\beta\sin\alpha\big)\bigg)$ \\
\hline
$AW^{+}_{\mu}W^{'-}_{\nu}$ & $\frac{ie^{2}v\sin2\beta}{4x\sin^{2}\theta_{w}}\bigg(1 +\frac{x^{2}}{4}\bigg)$ \\
\hline
$HZ_{\mu}Z_{\nu}$ & $\frac{e^{2}v}{16\sin^{2}\theta_{w}\cos^{2}\theta_{w}}\bigg(\sqrt{2}\cos\beta\sin\alpha + 4\cos\alpha\sin\beta\bigg)$ \\
\hline
$hZ_{\mu}Z_{\nu}$ & $\frac{e^{2}v}{16\sin^{2}\theta_{w}\cos^{2}\theta_{w}}\bigg(\sqrt{2}\cos\beta\cos\alpha - 4\sin\alpha\sin\beta\bigg)$ \\
\hline
$hZ^{'}_{\mu}Z^{'}_{\nu}$ & $\frac{e^{2}v\cos\alpha\cos\beta}{2\sqrt{2}x^{2}\sin^{2}\theta_{w}}\bigg(1 + \frac{x^{2}}{8}\big(1 - 5\cos2\theta_{w}\big)\sec^{2}\theta_{w}\bigg)$ \\
\hline
$HZ^{'}_{\mu}Z^{'}_{\nu}$ & $\frac{e^{2}v\cos\beta\sin\alpha}{2\sqrt{2}x^{2}\sin^{2}\theta_{w}}\bigg(1 + \frac{x^{2}}{8}\big(1 - 5\cos2\theta_{w}\big)\sec^{2}\theta_{w}\bigg)$ \\
\hline
$hZ_{\mu}Z^{'}_{\nu}$ & $\frac{e^{2}v\sec^{3}\theta_{w}}{8\sqrt{2}x\sin^{2}\theta_{w}}\cos\alpha\cos\beta\bigg(5 - 2\cos2\theta_{w} + \cos4\theta_{w}\bigg)$ \\
\hline
$HZ_{\mu}Z^{'}_{\nu}$ & $\frac{e^{2}v\sec^{3}\theta_{w}}{8\sqrt{2}x\sin^{2}\theta_{w}}\sin\alpha\cos\beta\bigg(5 - 2\cos2\theta_{w} + \cos4\theta_{w}\bigg)$ \\
\hline
$H^{-}W^{+}_{\mu}Z_{\nu}$  & $ - \frac{ie^{2}x^{2}v}{32\cos^{3}\theta_{w}}\cos\beta\sin\beta$ \\
\hline
$H^{-}W^{'+}_{\mu}Z_{\nu}$ & $- \frac{ie^{2}v\sin2\beta\sec\theta_{w}}{4x\sin^{2}\theta_{w}}\bigg(1 + \frac{x^{2}}{4}\bigg)$ \\
\hline
$H^{-}W^{+}_{\mu}Z^{'}_{\nu}$ & $\frac{ie^{2}v\sin2\beta}{4x\sin^{2}\theta_{w}}\bigg(1 + \frac{x^{2}}{16}\big(3 + \cos2\theta_{w}\big)\sec^{2}\theta_{w}\bigg)$\\
\hline
$H^{-}W^{'+}_{\mu}Z^{'}_{\nu}$ & $-\frac{ie^{2}v\tan^{2}\theta_{w}\sin2\beta}{8\sin^{2}\theta_{w}}\bigg(1 + \frac{x^{2}}{4}\sec^{2}\theta_{w}\bigg)$ \\
\hline
\end{tabular}
}
\caption{All the triple point $VVX$ couplings in the models where $V$ is a (light or heavy) gauge boson and $X$ is a scalar. These have been computed to $\mathcal{O}(x^2)$.}
\label{tab:VVX}
\end{center}
\end{table}
A couple of comments are in order.
\begin{enumerate}
\item \textbf{The Alignment Limit}: The $hWW$ coupling is changed from the corresponding SM one by a rather unwieldy factor as opposed to the 2HDM where the scaling factor is neatly given by $\sin(\beta-\alpha)$. This is due to the extended nature of EWSB in the model, as the $\Sigma$ model field that breaks the $SU(2)\times SU(2)$ down to a diagonal $SU(2)_L$ also feeds into the mechanism. Nevertheless, the scaling is still controlled by the two parameters $\alpha$ and $\beta$. The limit in which one extracts a SM-like higgs boson, i.e., the decoupling limit, would thus not be the same as in the 2HDM. In Fig.~\ref{fig:align} below, we show this alignment limit in the $\tan\beta-\sin(\beta-\alpha)$ plane commonly employed in the 2HDM literature. It can be seen that identifying the lighter mass eigenstate as the SM-like Higgs and demanding that its coupling to $WW$ take on the SM value forces the 2HDM to be confined to the regions very close to $\sin(\beta-\alpha)=\pm 1$, while in this model there is a fairly larger parameter space that opens up in the region $-0.4\leq\sin(\beta-\alpha)\leq 0.3$ for all values of $\tan\beta$\footnote{Interestingly the region around $\sin(\beta-\alpha)=0$ is the alignment limit in the 2HDM corresponding to identifying the heavier $H$ as the SM-like Higgs. See also Footnote [7].}.

\begin{center} 
 \begin{figure}[h!]
\includegraphics[scale = 0.4]{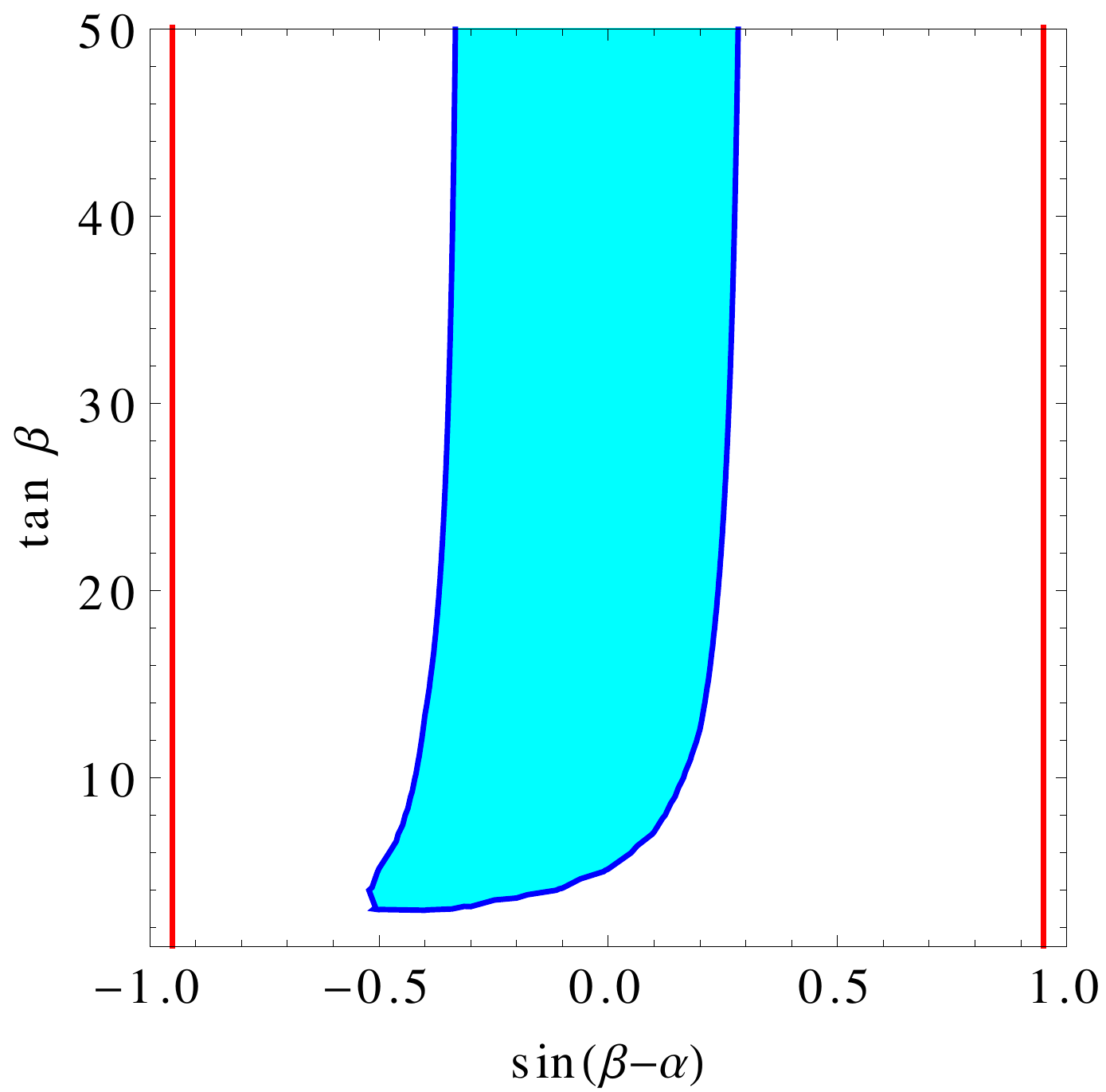}
\label{fig:align}
\caption{The alignment limit in the 2HDM (shown in red) and in this model (shown as the shaded region within the blue contour). As can be seen, while the 2HDM alignment limit is forced near the $\sin(\beta-\alpha)\approx\pm 1$ regions, the corresponding limit for this model occurs in the region $-0.5\leq\sin(\beta-\alpha)<0.3$ owing to the nature of the $g_{hWW}$ coupling. }
\end{figure}
\end{center}
\item \textbf{Perturbativity}: Couplings involving one (two) heavy gauge bosons are proportional to $1/x$ ($1/x^2$) - treating $x$ as a small parameter as we have done, we note that these couplings become large for sufficiently small $x$ (or equivalently, a sufficiently large $m_{W'}$). Thus, novel decays involving these heavy gauge bosons need to be treated with care and are not valid for arbitrarily massive $W'$s as we would lose perturbativity. As can be seen from Tables~\ref{tab:charged} and \ref{tab:neutral}, a similar pattern also emerges in the coupling of a $W'$ or a $Z'$ with two heavy fermions.
\end{enumerate}
Similarly we present the $VXX$ couplings in Table \ref{tab:VXX} - again we see similar features emerge where couplings involving a heavy gauge boson are enhanced by a factor of $1/x$. The presence of these vertices makes the phenomenology of this model rich - in addition to the usual decay channels for the heavy Higgs bosons ($WW$, $ZZ$, $bb/\tau\tau$) etc., there exist possibilities of them decaying into one of these heavy gauge bosons that would potentially lead to stark signals at the LHC involving, for example, multi-lepton final states. Since these heavy gauge bosons do not couple to the SM fermions, they evade much of the direct bounds and the oblique ones and thus need not be terribly massive. We discuss these possibilities further in Section~\ref{sec:pheno}.

\begin{table}[h!]
\begin{center}
\resizebox{\textwidth}{!}
{
\begin{tabular}{|   c   ||   c   |}
\hline
Vertex & Strength \\
\hline \hline
$A_{\mu}H^{+}H^{-}$ & $e\big(p_{H^{+}} - p_{H^{-}}\big)_{\mu}$ \\
\hline
$Z_{\mu}H^{+}H^{-}$ & $\frac{e\cos2\theta_{w}}{2\sin\theta_{w}\cos\theta_{w}}\bigg(1 + \frac{x^{2}\sec^{2}\theta_{w}}{128}\big(12 + 8\cos2\beta + 2\cos2(\beta - \theta_{w}) + 2\cos2(\beta + \theta_{w}) + 3\cos2\theta_{w} + 4\cos4\theta_{w} + \cos6\theta_{w}\big)\bigg)\big(p_{H^{-}} - p_{H^{+}}\big)_{\mu}$ \\
\hline
$Z^{'}_{\mu}H^{+}H^{-}$ & $\frac{e\sec^{2}\theta_{w}}{2x\sin\theta_{w}}\bigg(\sin^{2}\beta - \frac{x^{2}}{64}\big(30 + 2\cos2\beta - \cos2(\beta - \theta_{w}) + 2\cos2\theta_{w} - \cos2(\beta + \theta_{w}) \big)\bigg)\big(p_{H^{-}} - p_{H^{+}}\big)_{\mu}$ \\
\hline
$W^{+}_{\mu}H^{-}A$ & $\frac{e}{2\sin\theta_{w}}\bigg(1 + \frac{x^{2}}{32}\big(5 + 3\cos2\beta\big)\bigg)\big(p_{H^{-}} - p_{A}\big)_{\mu}$ \\ 
\hline
$W^{'+}_{\mu}H^{-}A$ & $\frac{e}{2x\sin\theta_{w}}\bigg(\sin^{2}\beta - \frac{x^{2}}{2}\bigg)\big(p_{H^{-}} - p_{A}\big)_{\mu}$ \\
\hline
$W^{+}_{\mu}H^{-}h$ & $\frac{ie\sin\alpha}{8\sin\theta_{w}}\bigg(\big(4\cos\beta + \sqrt{2}\sin\beta\big) + \frac{x^{2}}{8}\big(8\cos\beta - \sqrt{2}\sin\beta\big)\bigg)\big(p_{H^{-}} - p_{h}\big)_{\mu}$ \\
\hline
$W^{+}_{\mu}H^{-}H$ & $- \frac{ie}{8\sin\theta_{w}}\bigg(\big(4\cos\alpha\cos\beta - \sqrt{2}\sin\alpha\sin\beta\big) + \frac{x^{2}}{8}\big(8\cos\alpha\cos\beta + \sqrt{2}\sin\alpha\sin\beta\big)\bigg)\big(p_{H^{-}} - p_{H}\big)_{\mu}$ \\
\hline
$W^{'+}_{\mu}H^{-}H$ & $\frac{ie}{2\sqrt{2}x\sin\theta_{w}}\bigg(\sin\alpha\sin\beta + \frac{x^{2}}{4\sqrt{2}}\big(4\cos\alpha\cos\beta - \sqrt{2}\sin\alpha\sin\beta\big)\bigg)\big(p_{H^{-}} - p_{H}\big)_{\mu}$ \\
\hline
$W^{'+}_{\mu}H^{-}h$ & $\frac{ie}{2\sqrt{2}x\sin\theta_{w}}\bigg(\sin\alpha\sin\beta - \frac{x^{2}}{4\sqrt{2}}\big(4\sin\alpha\cos\beta + \sqrt{2}\sin\alpha\sin\beta\big)\bigg)\big(p_{H^{-}} - p_{h}\big)_{\mu}$ \\
\hline
\end{tabular}
}
\caption{All the triple point $VXX$ couplings in the models where $V$ is a (light or heavy) gauge boson and $X$ is a scalar. These have been computed to $\mathcal{O}(x^2)$.}
\label{tab:VXX}
\end{center}
\end{table}

We now turn to the Higgs-Fermion couplings - note that due to mixing, in addition to terms like $h\bar{f}f$ and $H\bar{f}f$, there would also be scalar couplings of the form $h\bar{f}F$ and $H\bar{f}F$. We list these couplings below working in the limit $\epsilon_R\to 0$ and to $\mathcal{O}(x^2)$. Notice also that for all light fermions $a_{u,d}\propto m \approx 0$ - however we have retained the $a$ terms below as these are important for the top sector (the generational index on the $\lambda$'s have been suppressed for brevity).

\begin{table}[h!]
\begin{center}
\renewcommand{\arraystretch}{2.1}
\resizebox{13cm}{!}
{
\begin{tabular}{|   c   ||   c | c | c | c|}
\hline
 & $\bar{u}_{L}u_{R}$ & $\bar{u}_{L}U_{R}$ & $\bar{U}_{L}u_{R}$ & $\bar{U}_{L}U_{R}$ \\
\hline \hline
$h$ &  $-\frac{\lambda_{u0}\sin\alpha}{\sqrt{2}}$ & $\frac{\lambda_{u0}\sin\alpha}{2} \frac{a_{u}x} {(a^{2}_{u} - 1)} $ & $- \frac{\lambda_{u0}\sin\alpha}{2} \frac{x }{a^{2}_{u} - 1}$ & $- \frac{\lambda_{u0}\sin\alpha}{2\sqrt{2}}\frac{a_u x^2 }{a^{2}_{u} - 1} $ \\ 
\hline
$H$ & $ \frac{\lambda_{u0}\cos\alpha}{\sqrt{2}} $ & $\frac{\lambda_{u0}\cos\alpha}{2}\frac{a_{u}x }{(a^{2}_{u} - 1)}$ & $\frac{\lambda_{u0}\cos\alpha}{2} \frac{x}{a^{2}_{u} - 1} $ & $\frac{\lambda_{u0}\cos\alpha}{2\sqrt{2}}\frac{a_u x^2 }{a^{2}_{u} - 1} $\\
\hline
$A$ & $- \frac{i\lambda_{u0}\cos\beta}{\sqrt{2}}$ & $\frac{i\lambda_{u0}\cos\beta}{2} \bigg( \frac{a_{u}x}{(a^{2}_{u} - 1)}  - \frac{x \tan^2\beta}{\sqrt{2}a_{u}} \bigg) $ & $-\frac{i\lambda_{u0}\cos\beta}{2} \frac{x}{a^{2}_{u} - 1}$ & $- \frac{i\lambda_{u0}\cos\beta}{2}\frac{x}{a^{2}_{u} - 1} $ \\
\hline
\end{tabular}
}
\caption{The couplings of the neutral scalars to the fermions in the model - these have been evaluated in the limit $\epsilon_R\to 0$ and to $\mathcal{O}(x^2)$. The couplings with the down-type fermions can be obtained via the substitutions $\lambda_{u0}\to \lambda_{d0}$ and $a_u \to a_d$. }
\end{center}
\end{table}
To leading order, $\lambda_{u0}=\frac{\sqrt{2}m_u}{v\sin\beta}$ (see Eqn.~\ref{eq:fermionmass}). Thus, it can be seen that the light fermion couplings to the neutral scalars follows a Type I 2HDM-like pattern with the scaling factor (relative to the SM) given by\footnote{In the Type I 2HDM, the $h (H)$ scaling has the $\cos\alpha \,(\sin\alpha)$ - this can, however, be accounted for by a redefinition of the mixing angle $\alpha$. It is identical to Type I 2HDM in the sense that the scaling does not differ between the up-type and down-type fermions. }
\begin{align}
\begin{split}
\xi_h^{u,d}&=\sin\alpha/\cos\beta, \\ 
\xi_H^{u,d}&=\cos\alpha/\cos\beta, \\ 
\xi_A^{u,d}&=\cot\beta.
\end{split}
\end{align}

In addition, the charged Higgs couplings to the light fermions are given (to leading order) by the following expressions:
\begin{align}
\begin{split}
H^{+}\bar{u}_{R}d_{L} + h.c. &= i\lambda^{ij}_{u0}\cos\beta, \\ 
H^{+}\bar{u}_{L}d_{R} + h.c. &= - i\lambda^{ij}_{d0}\cos\beta.
\end{split}
\end{align}
\subsection{Vacuum Stability}

Let us begin by writing the potential of the scalar fields again:
\begin{align*}
V(\Phi_{1},\Phi_{2},\Sigma) & = m^{2}_{11}\Phi^{\dagger}_{1}\Phi_{1} + m^{2}_{22}\Phi^{\dagger}_{2}\Phi_{2} - (m^{2}_{12}\Phi^{\dagger}_{1}\Sigma\Phi_{2} + \textnormal{h.c.}) + \frac{\beta_{1}}{2}(\Phi^{\dagger}_{1}\Phi_{1})^{2} + \frac{\beta_{2}}{2}(\Phi^{\dagger}_{2}\Phi_{2})^{2} \\
& + \beta_{3}(\Phi^{\dagger}_{1}\Phi_{1})(\Phi^{\dagger}_{2}\Phi_{2}) + \beta_{4}(\Phi^{\dagger}_{1}\Sigma\Phi_{2})(\Phi^{\dagger}_{2}\Sigma^{\dagger}\Phi_{1}) + \bigg[\frac{\beta_{5}}{2}(\Phi^{\dagger}_{1}\Sigma\Phi_{2})^{2} + \textnormal{h.c.}\bigg].
\end{align*}
Defining $\Phi^{\dagger}_{1}\Phi_{1} = a$, $\Phi^{\dagger}_{2}\Phi_{2} = b$, $\textnormal{Re}[\Phi^{\dagger}_{1}\Sigma\Phi_{2}] = c$ and $\textnormal{Im}[\Phi^{\dagger}_{1}\Sigma\Phi_{2}] = d$, the quartic part of the Lagrangian (which needs to be a manifestly positive quantity so the potential can be bounded from below) can now be recast in the following form:
\begin{align}
\begin{split}
V_{4} &= \frac{1}{2}\big(\sqrt{\beta_{1}}a - \sqrt{\beta_{2}}b\big)^{2} + \big(\beta_{3} + \sqrt{\beta_{1}\beta_{2}}\big)(ab - c^{2} - d^{2}) + 2(\beta_{3} + \beta_{4} + \sqrt{\beta_{1}\beta_{2}})c^{2} \\
& + \big(\textnormal{Re}[\beta_{5}] - \beta_{3} - \beta_{4} - \sqrt{\beta_{1}\beta_{2}}\big)(c^{2} - d^{2}) - 2cd\, \textnormal{Im}[\beta_{5}].
\end{split}
\end{align}
With the field identifications $\Phi_1\to \Phi_1$ and $\Phi_2\to \Sigma\Phi_2$, this is identical to the corresponding potential in the 2HDM literature - see, for instance \cite{Bhattacharyya:2015nca}. We can now write down the (Cauchy-Schwarz) inequality for the combination of fields in the following form:

\be
ab \geq c^{2} + d^{2}.
\ee
By considering various choices of field directions and ensuring that $V_4$ does not run negative for large values of the parameters $a,b$ leads us to the following constraints on the various $\beta$'s:
\begin{align}
\begin{split}
&\beta_{1} > 0 \\
&\beta_{2} >0 \\
&(\beta_{3} + \sqrt{\beta_{1}\beta_{2}}) > 0 \\
&\beta_{3} + \beta_{4} + \sqrt{\beta_{1}\beta_{2}} > |\beta_{5}|.
\end{split}
\end{align}

Translated into the space of the $\lambda$'s as given in Eqn.~\ref{eqn:L1}, these constraints read

\begin{align}
\begin{split}
&\lambda_{1} + \lambda_{3} > 0 \\ 
&\lambda_{2} + \lambda_{3}  > 0 \\
&2\lambda_{3} + \lambda_{4} + 2\sqrt{(\lambda_{1} + \lambda_{3})(\lambda_{2} + \lambda_{3})} > 0 \\
&2\lambda_{3} + \frac{1}{2}(\lambda_{5} + \lambda_{6}) - \frac{1}{2}|\lambda_{5} - \lambda_{6}| + 2\sqrt{(\lambda_{1} + \lambda_{3})(\lambda_{2} + \lambda_{3})} > 0.
\end{split}
\end{align}
These conditions guarantee that the electroweak vacuum is stable in this model - the results are formally identical to that found in the 2HDM literature. An extensive discussion of the stability bounds on the 2HDM potential can be found in \cite{Klimenko:1984qx,Maniatis:2006fs,Nie:1998yn}.
\subsection{Unitarity}

\begin{center}
\begin{figure}[h!]
\includegraphics[scale=0.4]{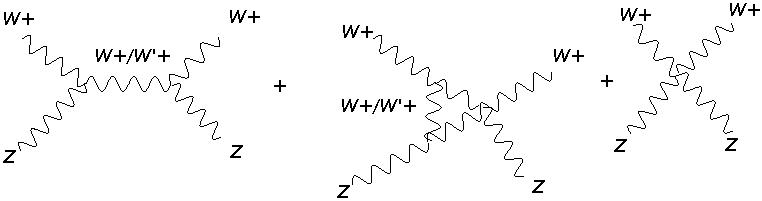}
\includegraphics[scale=0.4]{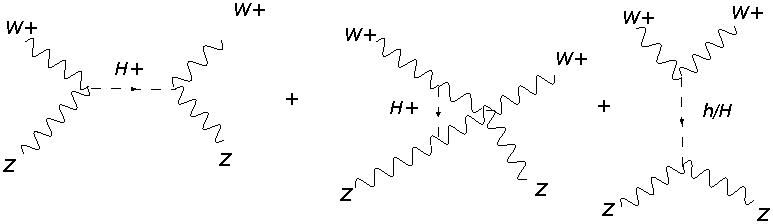}
\caption{Gauge Boson and Scalars mediated Feynman diagrams contributing to the scattering process $W^{+}Z \rightarrow W^{+}Z$.}
\label{fig:stuWW}
\end{figure}
\end{center}

\begin{center}
 \begin{figure}[h!]
\includegraphics[scale=0.4]{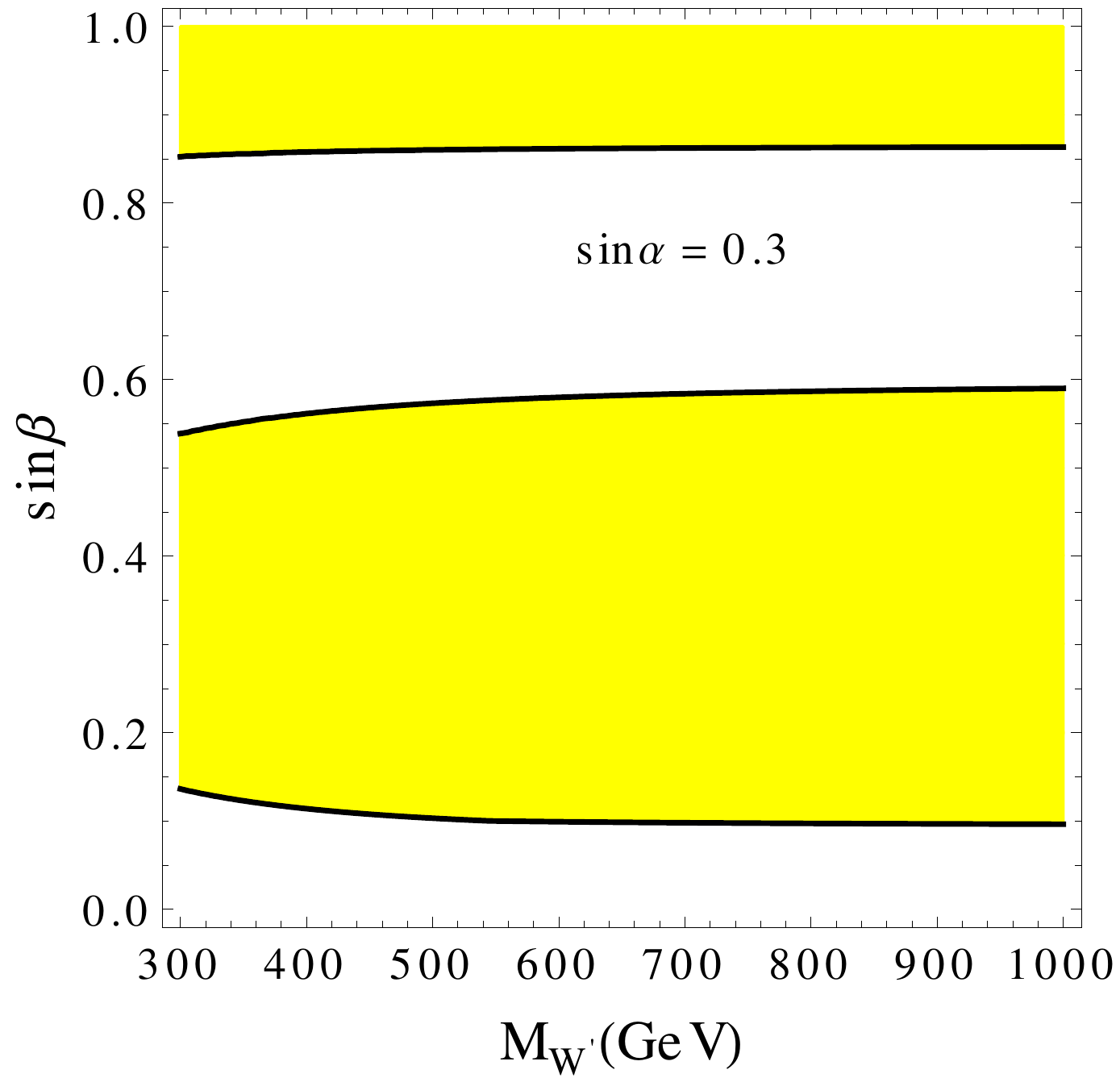}\hspace{0.1in}
\includegraphics[scale=0.4]{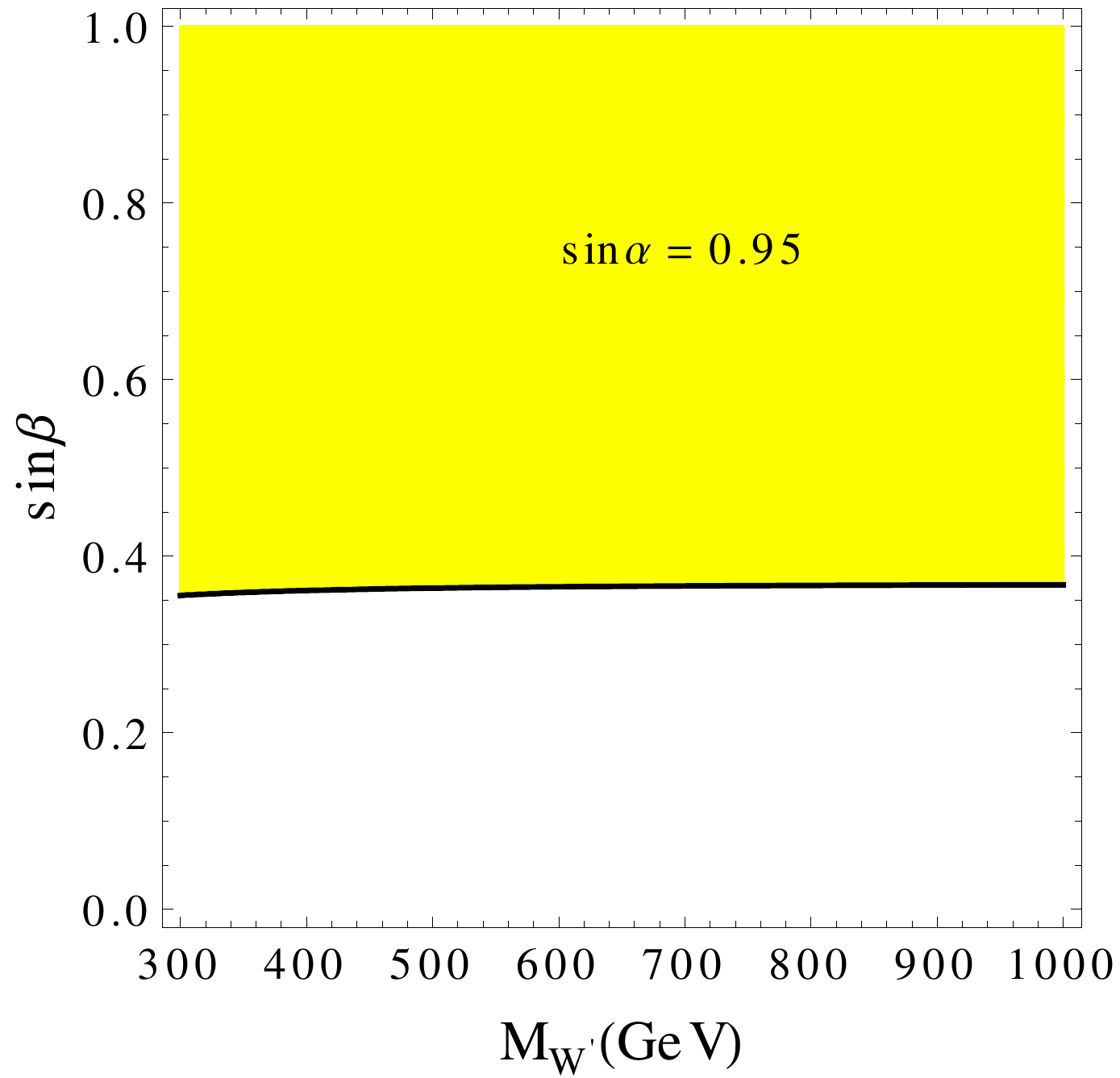}\hspace{0.1in}
\includegraphics[scale=0.4]{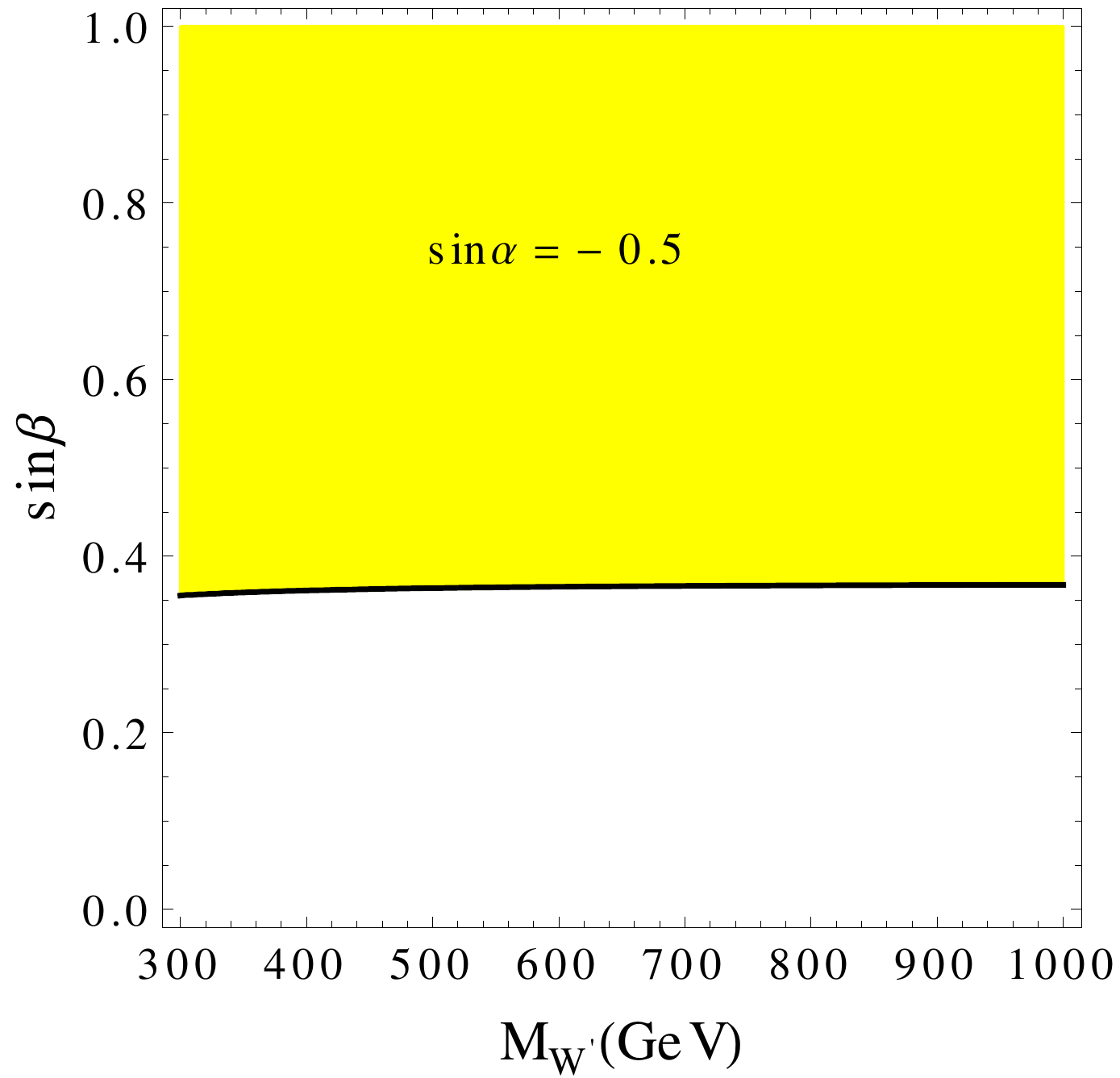}
\caption{The yellow shaded region corresponds to the allowed benchmark points after recasting the unitarity condition for our model considering different value of $\sin\alpha$.}
\label{fig:Unitarity}
\end{figure}
\end{center}

Insuring perturbative unitarity in the $2 \rightarrow 2$ longitudinal vector boson scattering is an important aspect of every BSM scenario. This amplitude can be written in the schematic form
\be
\mathcal{M}(W_LW_L\to W_LW_L)= a_0+a_1\frac{E^4}{M_W^4}+a_2\frac{E^2}{M_W^2},
\ee
with a constant piece $a_0$ and terms with grow with energy as $E^2$ and $E^4$. In the SM, the relation between the four point and the three-point vertices arising out of gauge invariance cancels the $E^4$ growth, while the cancelation of the $E^2$ piece requires the higgs exchange diagrams - a fact that was used to first constrain the mass of the higgs boson \cite{Lee:1977eg}. Before the discovery of the higgs, ``higgsless" models of electroweak symmetry breaking emerged that guaranteed perturbative unitarity using exchange of a tower of KK bosons in lieu of a higgs \cite{Chivukula:2003kq,Csaki:2003dt}. The present model has both higgs and heavy gauge bosons and so unitarization of the longitudinal vector boson scattering can proceed via both channels. We will analyse this for the process $W_{L} Z_{L} \rightarrow W_{L} Z_{L}$. Writing the amplitude in the form  
\be
\mathcal{M}(W^{\pm}_{L}Z_{L} \rightarrow W^{\pm}_{L}Z_{L} ) = \mathcal{M}_{4} + \mathcal{M}_{s} + \mathcal{M}_{t} + \mathcal{M}_{u}, \nonumber
\ee
we see that the amplitudes arise from all tree-level processes listed in Fig[~\ref{fig:stuWW}.  The $\mathcal{O}(E^4)$ growth is canceled by imposing the sum rule 
\begin{equation}
g_{WWZZ} = g^{2}_{WWZ} + g^{2}_{WW^{'}Z}.
\end{equation}
The first term is reminiscent of the SM wherein the relation between the four- and three-point couplings is guaranteed by gauge invariance while the second arises from the $W'$ exchange diagram.
Similarly, to cancel all the quadratic order energy growth, all the three-point vertices involving gauge bosons and Higgs bosons must  be constrained by

\begin{equation}
 \bigg(\frac{M^{4}_{Z}}{M^{2}_{W}}g^{2}_{WWZ} + \frac{M^{4}_{Z}}{M^{2}_{W^{'}}}g^{2}_{WW^{'}Z}\bigg) =  g_{HWW}g_{HZZ} + g_{hWW}g_{hZZ} - g^{2}_{H^{\pm}W^{\mp}Z}.  
\label{eq:UniWH}
\end{equation}
We note here that this involves both gauge and higgs mediated diagrams. In Fig.~\ref{fig:Unitarity} we show the allowed region in the $M_{W^{'}} - \sin\beta$ plane after imposing these sum rules for various choices of $\sin\alpha$.\footnote{The choice of the $\sin\alpha$ values anticipates Fig.~\ref{fig:h125}, wherein we have put in the LHC constraints on the $h$-125 higgs; $\sin\alpha=0.95$ and $-0.5$ are allowed, while $\sin\alpha=0.3$ is not.} 

\section{LHC Phenomenology}
\label{sec:pheno}
Finally with all the couplings and other relevant model details in place, we turn to the phenomenological issue of production and discovery of the various particles in the model. In this section, we concentrate on the scalar sector of the model and detail the cross-sections and branching ratios of the various higgses.\footnote{See Ref.~\cite{Chen:2019pkq} for a recent analysis in the Type-I 2HDM case.}We postpone a detailed collider analysis of this model including the gauge and the fermion sectors to a future work \cite{TTM2collider}.

\subsection{The 125 GeV Higgs}
As is true in any model with multiple CP-even scalars, the phenomenological analysis of the model will be different depending on which one is chosen to be SM-like. There are two choices in this model: choosing the lighter of the two mass eigenstates to be the SM-like Higgs boson will mean that we have another heavy Higgs while choosing the heavier mass eigenstate to be the 125 GeV Higgs would mean that there is another higgs \emph{lighter} than the SM one. We begin with an understanding of how the parameter spaces of the model will look like in the two distinct cases.

\begin{center}
 \begin{figure}[h!]
\includegraphics[scale = 0.4]{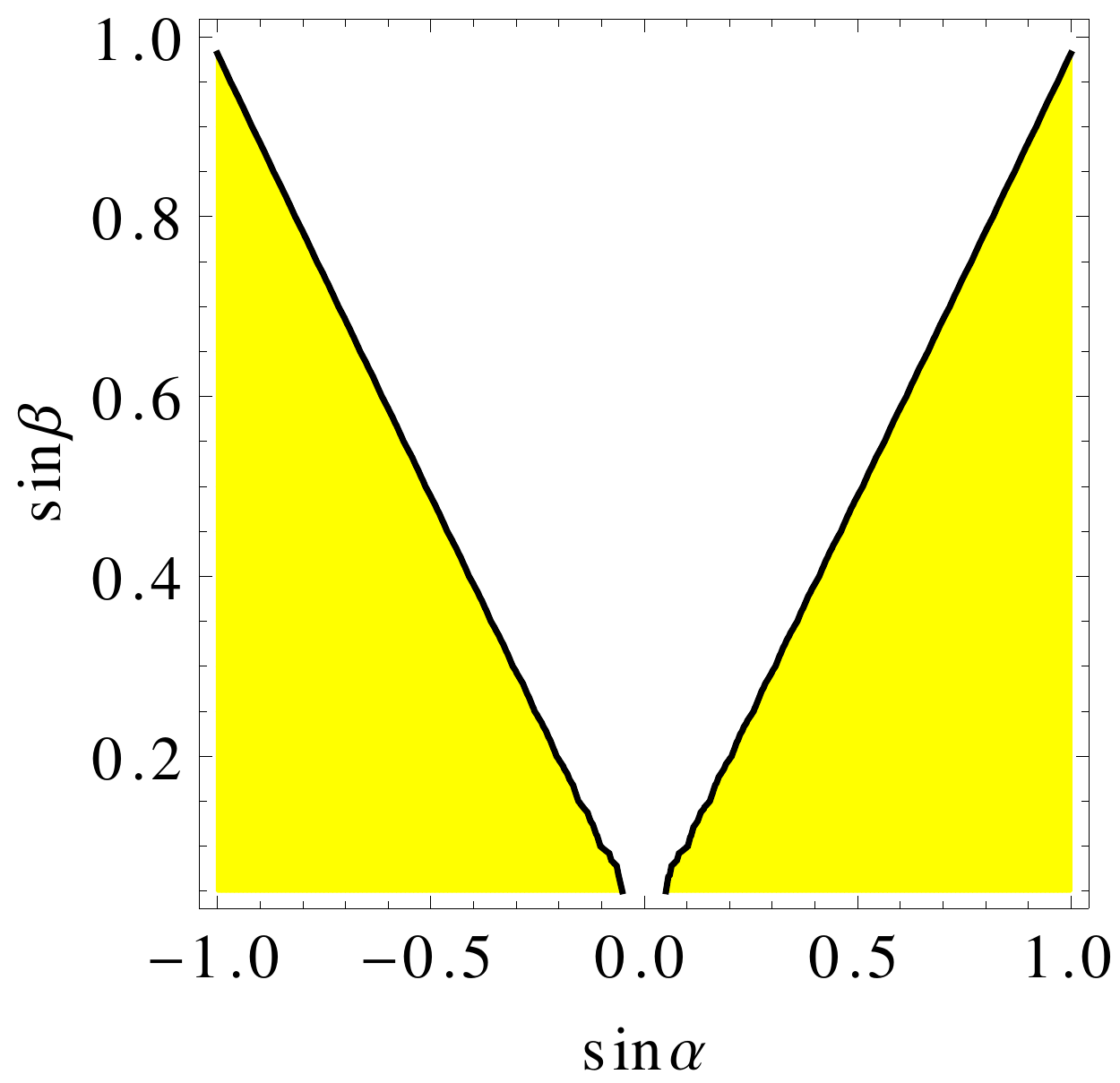}
\includegraphics[scale = 0.4]{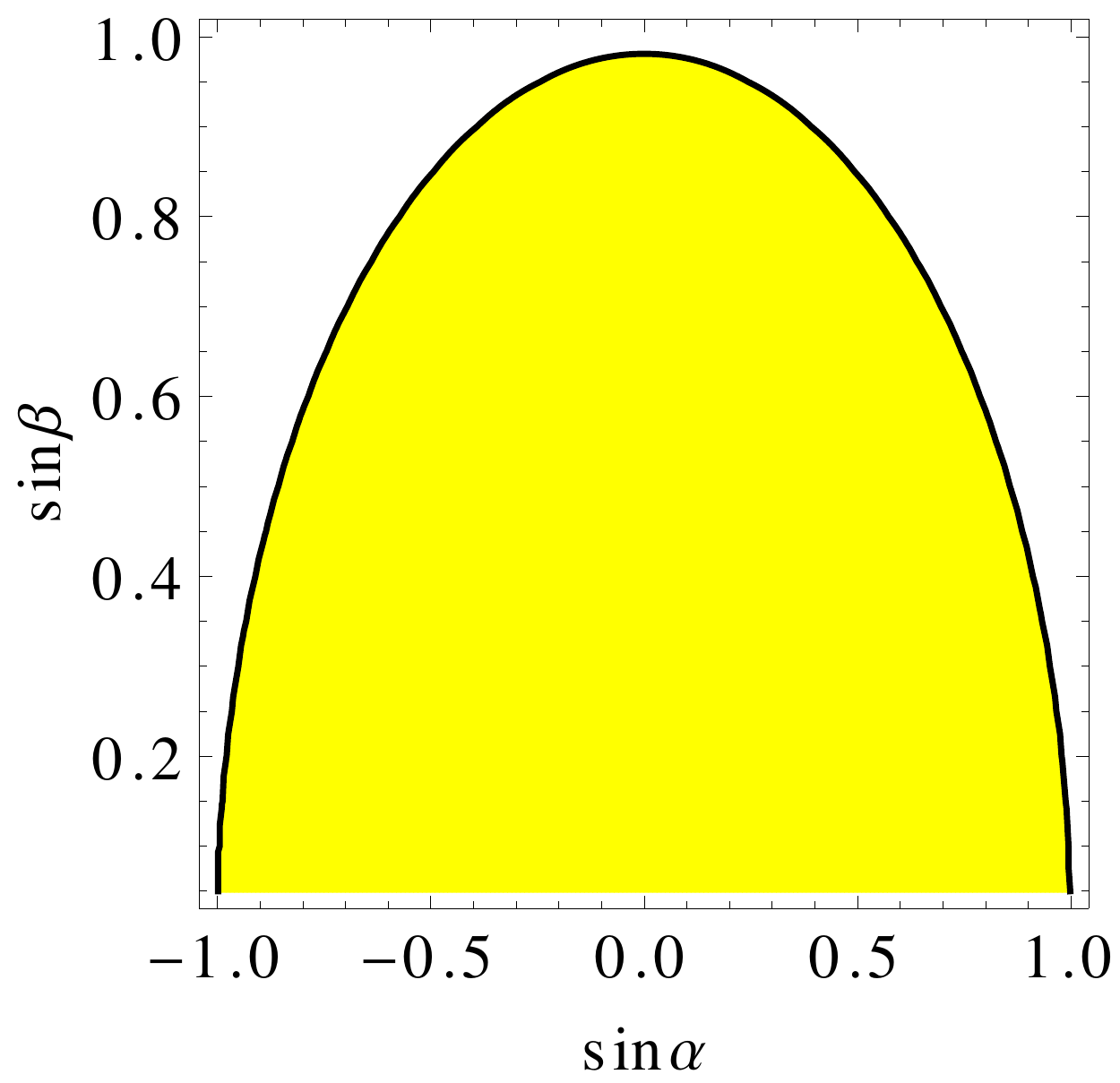}
\caption{The yellow shaded region corresponds to the 125 GeV lighter higgs $h$ (left) and the 125 GeV heavier higgs $H$ (right) yielding the reported value of gluon-fusion production cross-section.}
\label{fig:ggh}
\end{figure}
\end{center}

We first recall the scaling to be employed to facilitate the comparison to the SM case. Fixing, for instance, the lighter higgs to be SM-like, we have
\be
\frac{\sigma(gg \rightarrow h)}{\sigma(gg \rightarrow h_\textrm{SM})} = \frac{\Gamma(h \rightarrow gg)}{\Gamma(h_\textrm{SM} \rightarrow gg)}.
\label{eq:scaling1}
\ee
Since in the SM, the decay width of $h \rightarrow gg$ takes the form
\be 
\Gamma(h_\textrm{SM} \rightarrow gg) = \frac{\alpha^{2}m^{3}_{h}N^{2}_{C}}{256\pi^{3}v^{3}}|\sum_{i}e^{2}_{i}F_{i}|^{2},
\ee
we find
\be
\qquad \frac{\Gamma(h \rightarrow gg)}{\Gamma(h_\textrm{SM} \rightarrow gg)} = \frac{|\sum_{q_{i}, Q_{i}}\xi^{i}_{hff}e^{2}_{i}F_{i}|^{2}}{|\sum_{q_{i}}e^{2}_{i}F_{i}|^{2}},
\label{eq:scaling2}
\ee

where $F_{1/2} = - 2\tau\big[1 + (1 - \tau)f(\tau)\big]$, $\tau = \frac{4m^{2}_{f}}{m^{2}_{h}}$ and
\[
   f(\tau) = \left \{ \begin{array}{ll}
                 [\sin^{-1}\big(\sqrt{\tau^{-1}}\big)]^{2} &  \mbox{ $ \tau \geq 1 $} \\
                 - \frac{1}{4}\big|\ln\frac{1 + \sqrt{1 - \tau}}{1 - \sqrt{1 - \tau}} - i\pi \big|^{2}  & \mbox{$\tau <  1$} .
                 \end{array} \right. \]
 We have also defined the scaling factor $\xi_{hff} = \frac{g_{hff}}{g^{SM}_{hff}}$, and summed over both the SM and the heavy vector fermions in the loop. The latter are a non-negligible contribution because even as $M_D\to \infty$ and the heavy quarks formally decouple, the loop factor $F_{\frac{1}{2}}(\tau)$ contributes the asymptotic value $- \frac{4}{3}$. We use Eqns.~\ref{eq:scaling1} and \ref{eq:scaling2} to compute the rates in our model as simple multiples of the corresponding ones in the SM. For example, Fig.~\ref{fig:ggh} shows the range of parameter space available after imposing the reported SM value of the $gg\to h$ cross-section for the observed 125 GeV higgs in the $\sin\alpha-\sin\beta$ parameter space.\footnote{We note here that since the scaling of the higgs couplings in our model do not neatly factor as $\sin(\beta-\alpha)$, we choose a different parameter space from the one commonly employed in the 2HDM literature.} As can be seen readily, there is a wide range of the mixing angle $\alpha$ and the ratio $f/F$ that are admissible for both cases.
 
\begin{center}
\begin{figure}[h!]
\includegraphics[scale=0.4]{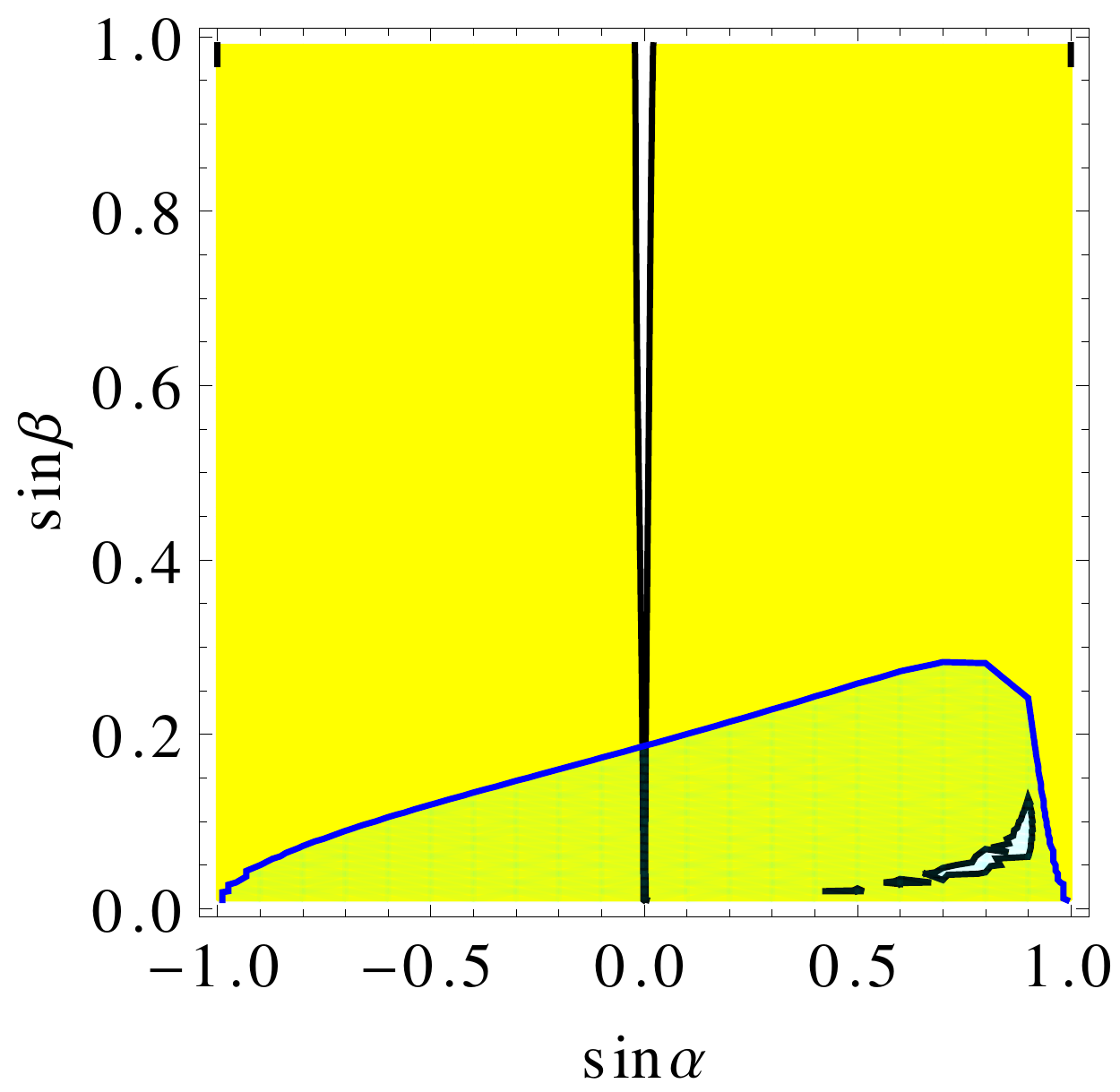}\hspace{0.1in}
\includegraphics[scale=0.4]{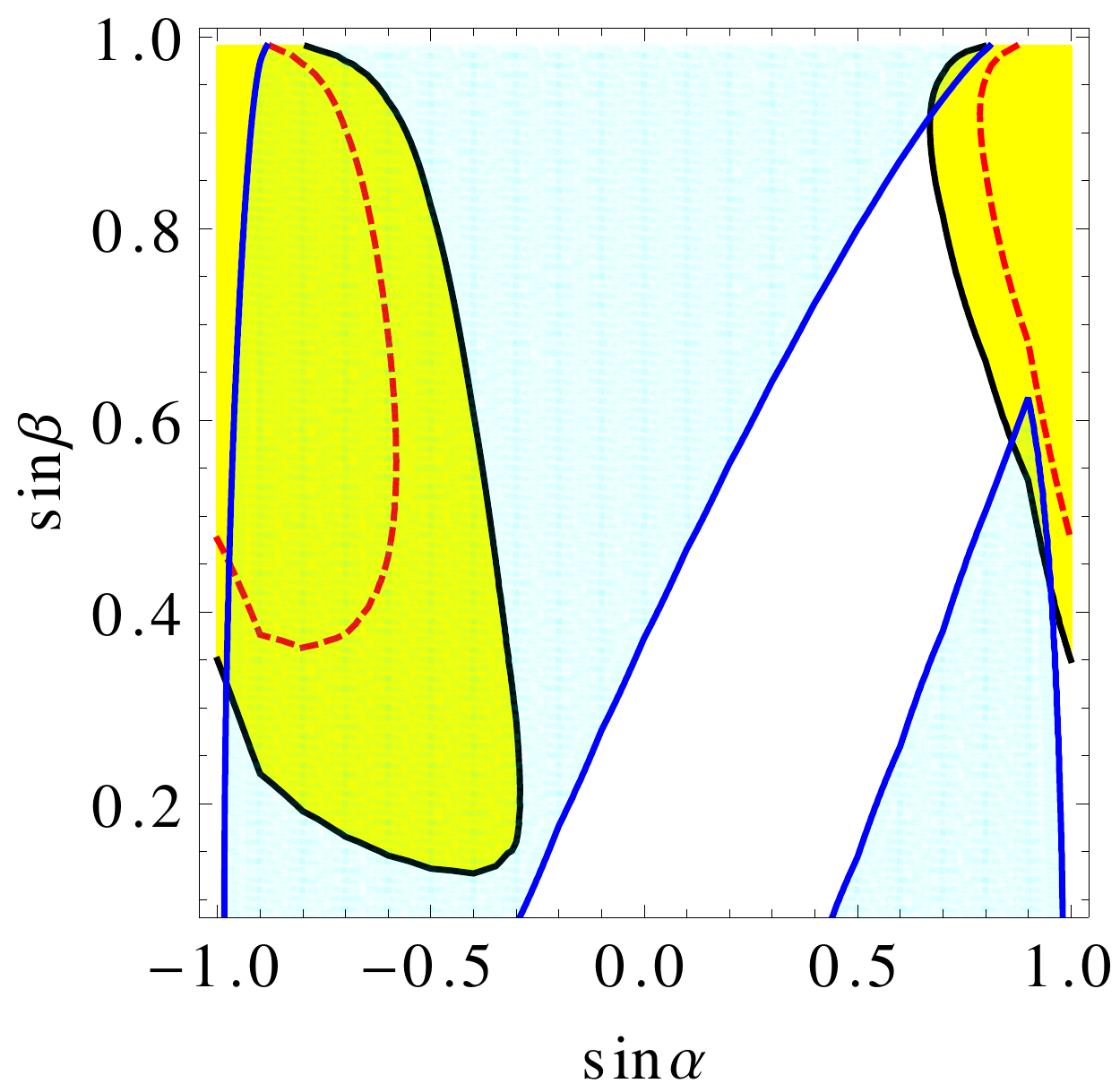}\hspace{0.1in}
\caption{The extent of parameter space admissible (regions in yellow) after demanding that the rates for the $g g \rightarrow \gamma \gamma$ (left), and the $g g \rightarrow h \rightarrow W W^{*}/Z Z^{*}$ (right) match the experimentally reported values for the 125 GeV higgs for the case of the light higgs $h$ being SM-like. Also overlaid in each plot are the corresponding regions in the Type I 2HDM (regions enclosed by the blue contours). The dashed red contours are the ATLAS $h\to b\bar{b}$ limits.}
\label{fig:h125}
\end{figure}
\end{center}

\begin{center}
 \begin{figure}[h!]
\includegraphics[scale=0.4]{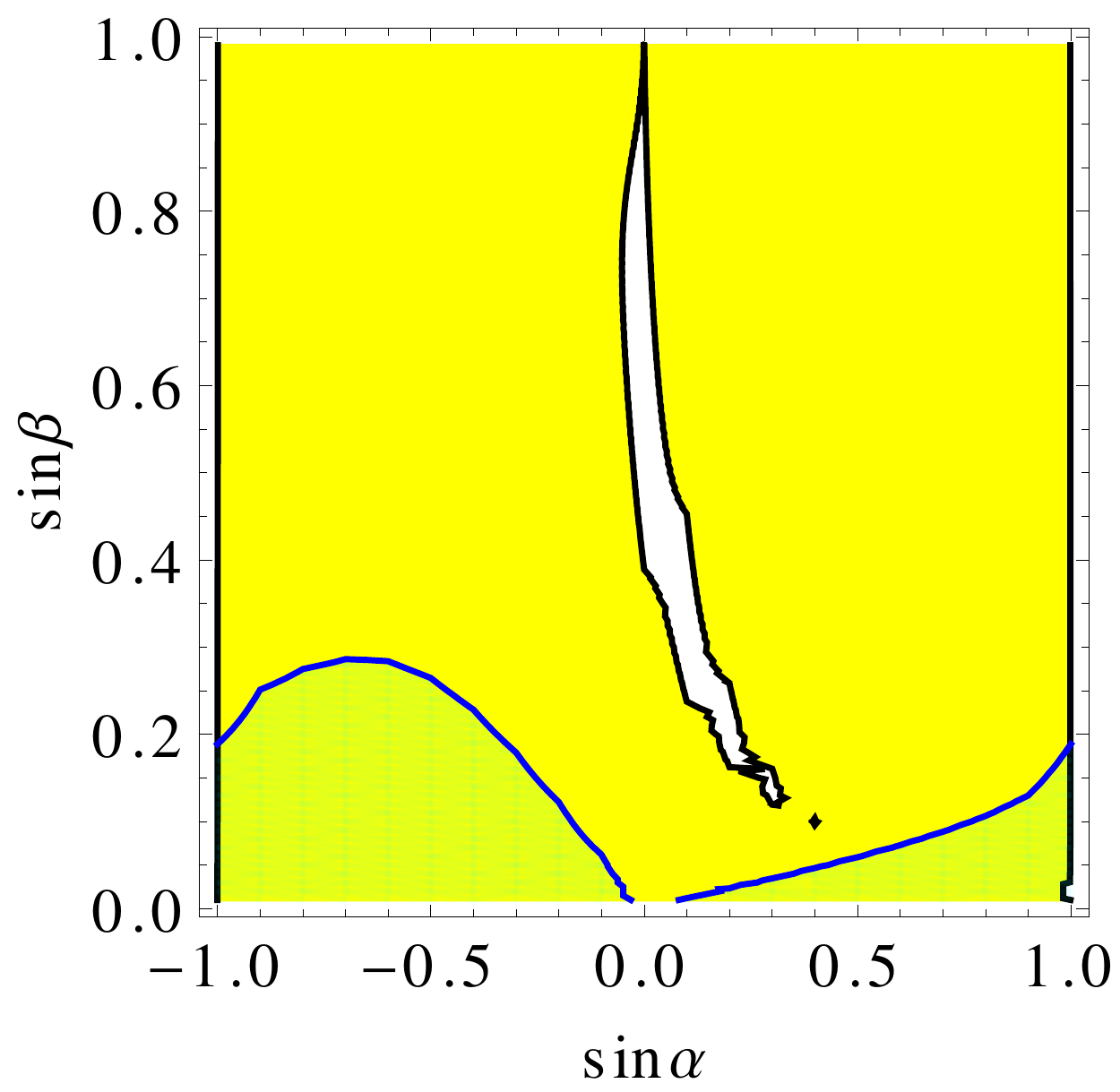}\hspace{0.1in}
\includegraphics[scale=0.4]{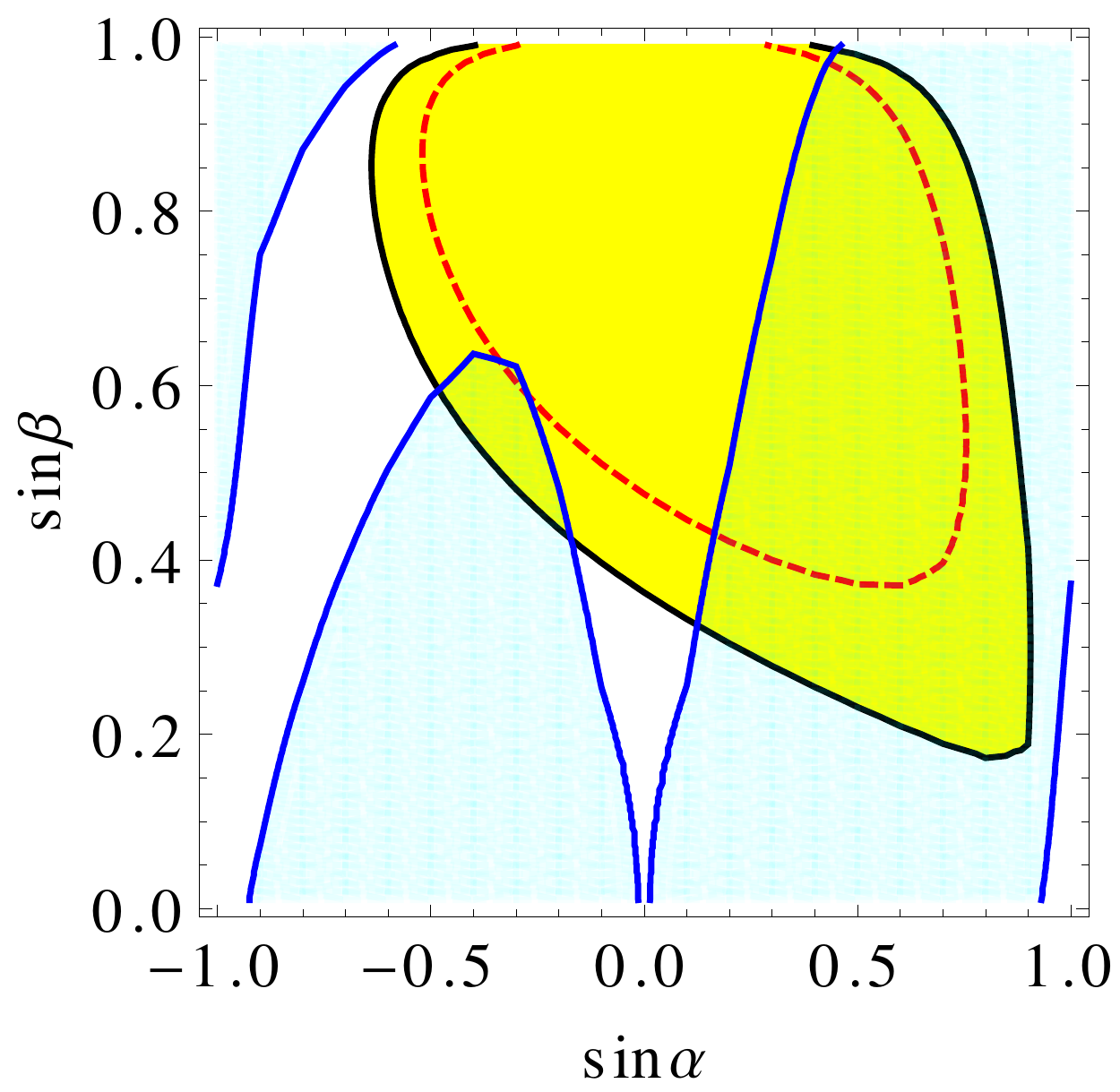}\hspace{0.1in}
\caption{The extent of parameter space admissible (regions in yellow) after demanding that the rates for the $g g \rightarrow \gamma \gamma$ (left), and the $g g \rightarrow h \rightarrow W W^{*}/Z Z^{*}$ (right) match the experimentally reported values for the 125 GeV higgs for the case of the heavy higgs $H$ being SM-like. Also overlaid in each plot are the corresponding regions in the Type I 2HDM (regions enclosed by the blue contours). The dashed red contours are the ATLAS $h\to b\bar{b}$ limits.}
\label{fig:H125}
\end{figure}
\end{center}

Of course, to make a realistic comparison, we need to compute $\sigma\times \textrm{BR}$ for the higgs in various channels and check that we consistently can reproduce the reported SM-higgs values for various choices of parameter values. A 125 GeV higgs primarily decays to $b\bar{b}$ ($\tau\bar{\tau}$), with small contributions from the off-shell channels like $WW^{*}$, $ZZ^{*}$ and finally the loop induced decay channels like $\gamma\gamma$, $Z\gamma$ and $gg$. In Fig.~\ref{fig:h125}, we show the regions of parameter space available after demanding that the $\sigma\times \textrm{BR}$ values corresponding to the $\gamma\gamma$, and the $WW^{*}/ZZ^{*}$ channel rates match the reported experimental value \cite{ATLAS:2020qdt,CMS:2020gsy} - superimposed on each plot is the corresponding region in Type I 2HDM.
It is interesting to observe that while there is a significant overlap between the two models, there are striking differences as well. Since there are additional particles that run in the loop - both heavy quarks (whose mass is set at $1.2$ TeV) and heavy gauge bosons (whose mass is set at 500 GeV), both the production rates and decay branching ratios are different from the 2HDM case. We see from Figs.~\ref{fig:h125} and \ref{fig:H125} that the $\gamma\gamma$ rate can be accommodated in almost the entire region of parameter space, while the $WW/ZZ$ case is more restricted regardless of which higgs is chosen to be SM-like. Type I 2HDM shows precisely the opposite behavior wherein the $WW/ZZ$ constraint is a little more relaxed in terms of the allowed parameter space. Putting in the LHC constraints, for the case of a light higgs being the 125 GeV particle, we find that the range $-0.3<\sin\alpha<-1$ covers almost the entire range of $\sin\beta$ values whereas for positive values of $\sin\alpha$, a smaller portion of admissible range opens up at $\sin\alpha\approx 0.7$ for values of $\sin\beta\approx 0.6$ and above. Both these regions shrink if we also impose the latest $h\to b\bar{b}$ (with the higgs produced via associated production) constraints \cite{ATLAS:2018nkp} - these are shown by the dashed red contours in Fig.~\ref{fig:h125}.

It is clear from the foregoing analysis that a simple measurement of rates in different channels alone would not serve to tell the models apart in spite of the present model having a richer EWSB structure. A precise determination of the decay widths of the higgs  would point to interesting physics beyond simple two-doublet extensions of the SM as it would reveal the $\mathcal{O}(x^2)$ corrections inherent in the couplings. Since $x\approx m_{W}/m_{W'}$, this would be more efficient if the $W'$ is not too massive - we remind the reader that in this model since there are no tree level couplings of the light fermions to the extra heavy gauge bosons,\footnote{The $Z'$ does have a small hypercharge coupling $\propto x$ to the SM fermions.} the heavy gauge bosons can comfortably evade all direct bounds coming from experiments. 

In Fig.~\ref{fig:H125} below, we perform a similar analysis fixing the heavier of the two eigenstates to be the SM-like higgs boson and compute the parameter space available after demanding SM-like rates. Comparing this to Fig.~\ref{fig:h125}, we see that the admissible parameter spaces in the two cases are largely exclusive of each other in spite of the production cross-sections having a significant overlap (see Fig.~\ref{fig:ggh}). 

\subsection{The Heavy Higgs $H$}

We now turn our attention to the task of identifying promising channels at the LHC that would uncover the particular new physics signals associated with this model. In all our calculations below, we fix $M_{D} = 1.2$ TeV, and $M_{W^{'\pm}} = 400$ GeV and work in the  $\sin\beta - \sin\alpha$ plane. While relaxing these would undoubtedly introduce more decay channels particularly for lighter $H$, in this section we restrict our attention to these benchmark points.

\begin{figure}[]
\includegraphics[scale=0.4]{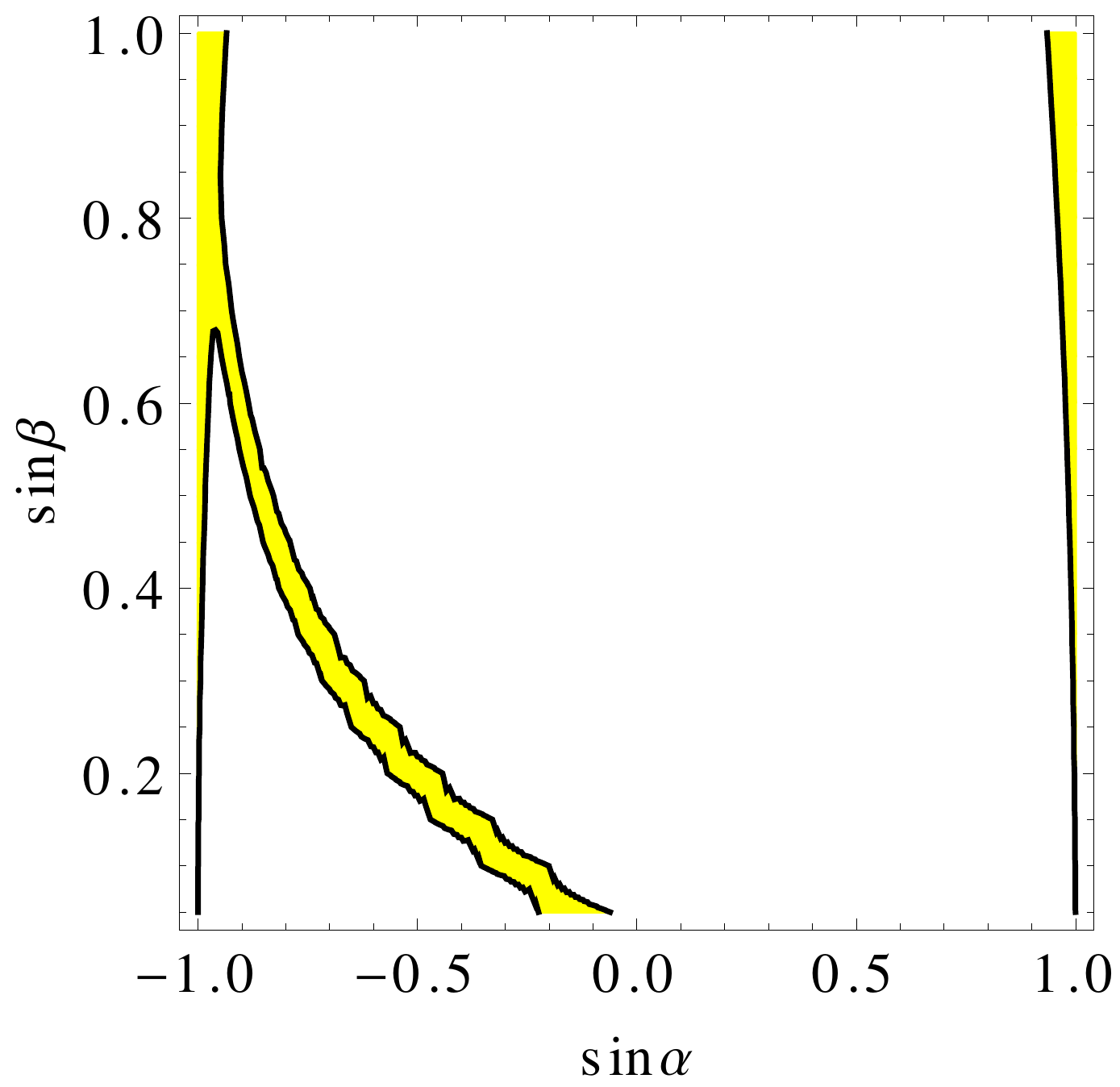}
\includegraphics[scale=0.4]{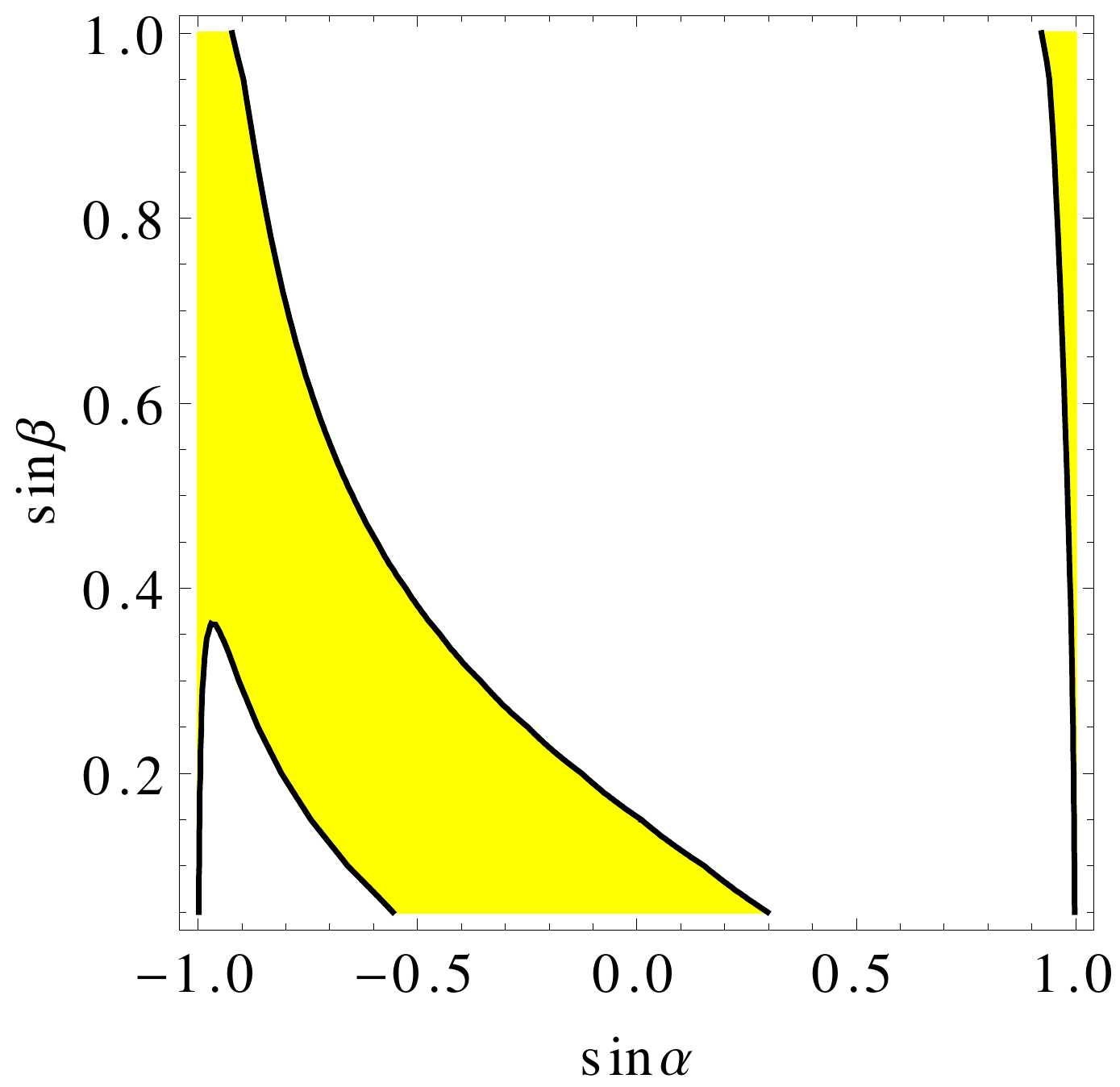}
\includegraphics[scale=0.4]{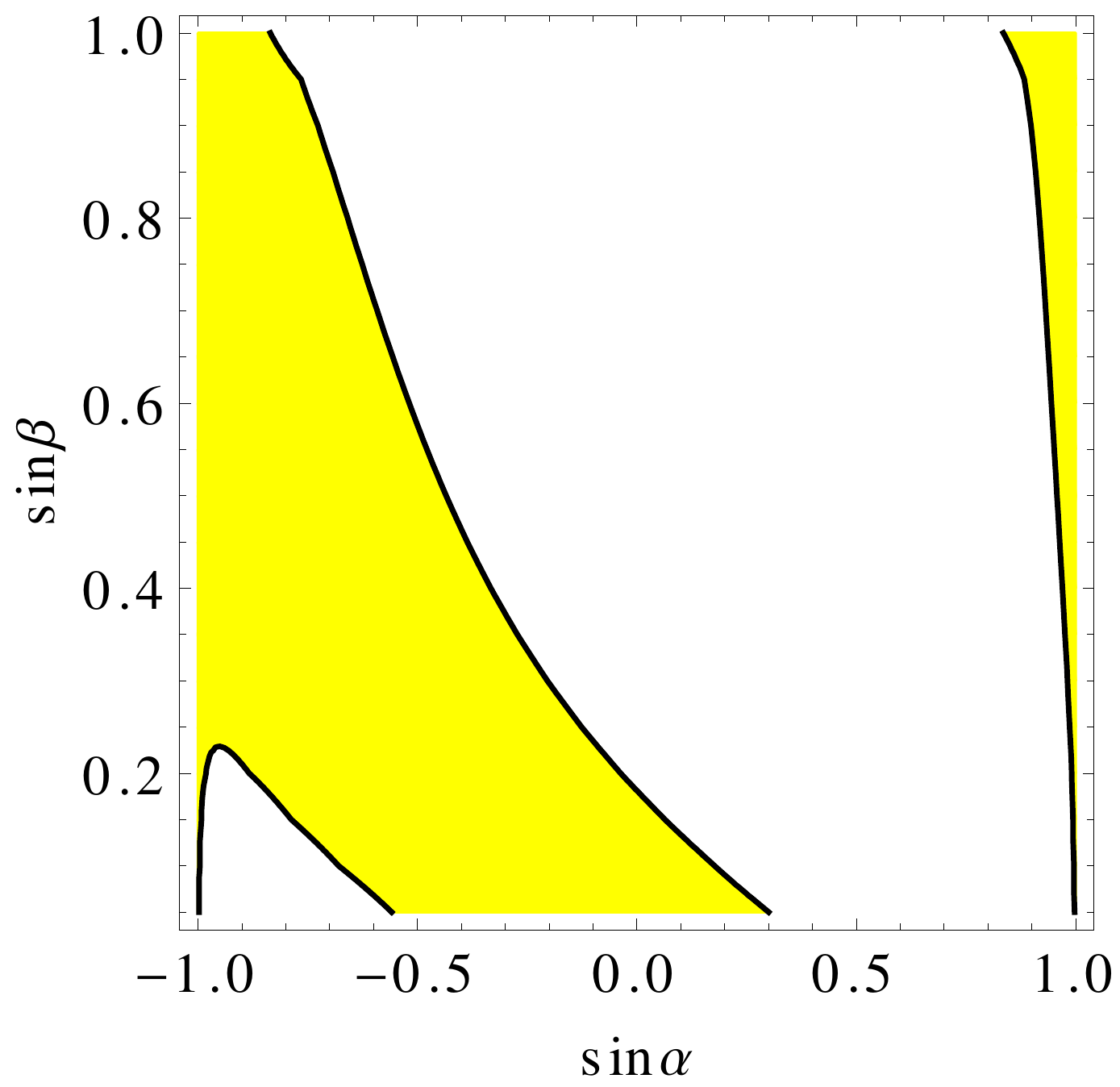}
\caption{The extent of parameter space admissible (regions in yellow) after imposing the cross-section bound from ATLAS in the $H\to ZZ$ channel for $M_H=$200 GeV (left), 600 GeV (middle), and 1 TeV (right). It is seen that the admissible region grows with increasing mass as the limit on the cross-section gets understandably weaker.}
\label{fig:Hlimit}
\end{figure}

\begin{center} 
 \begin{figure}[h!]
\includegraphics[scale = 0.8]{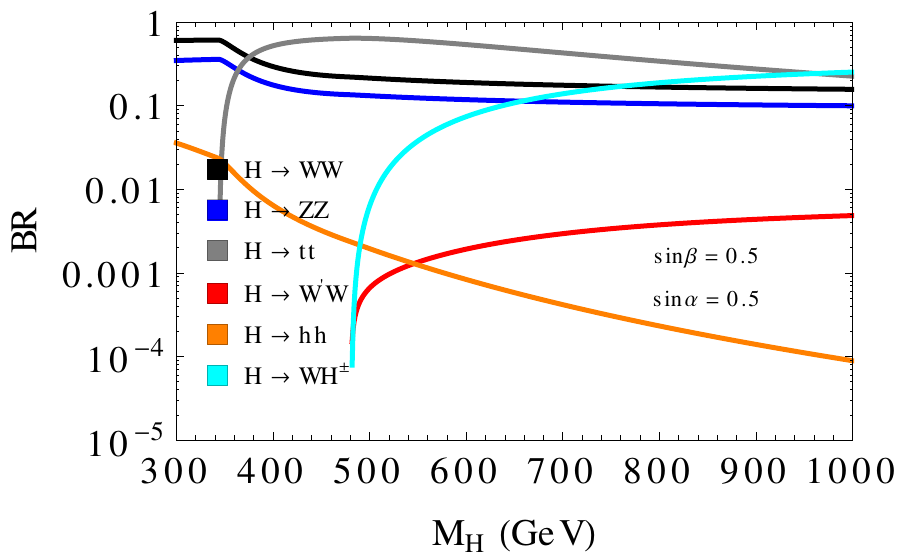}\hspace{0.1in}
\includegraphics[scale = 0.8]{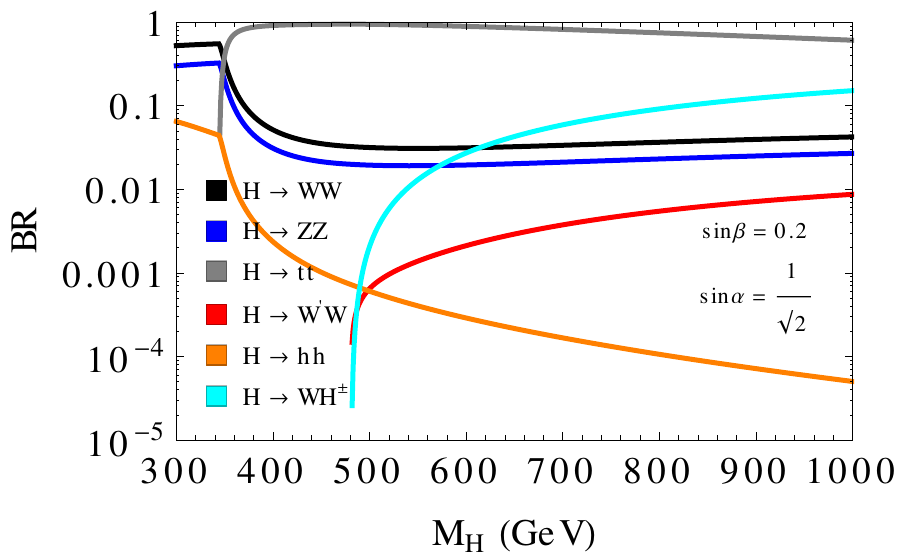}
\caption{The decay branching ratio of the heavy higgs $H$ into different final states for two different choices of $\sin\alpha$ and $\sin\beta$.}
\label{fig:HBR}
\end{figure}
\end{center}

\noindent The ATLAS and CMS experiments have looked for a heavy higgs in the diboson channel \cite{Aaboud:2017rel}, and have placed upper limits on the corresponding $\sigma\times$BR for the process for a wide range of heavy higgs masses -  in Fig.~\ref{fig:Hlimit}, we translate these limits to the case of a $H$ that is light ($M_H=200$ GeV), moderately heavy ($M_H=600$ GeV), and heavy  ($M_H=1000$ GeV). It is seen that the limits are stronger on lighter higgses, and that the allowed region saturates in a band around $-1\leq \sin\alpha \leq 0.3$ and a thin strip around $\sin\alpha=1$. It is seen from Fig.~\ref{fig:h125} that these regions also correspond to the ones that are allowed by the h-125 data - thus there is potential in this model for a heavier higgs to be discovered for a wide range of masses. We then turn to the question of understanding the best channels to do so.
We start by listing out the decay widths in all the available channels: in addition to the usual $VV$ and $ff$, there also exist decays into the lighter higgses, and $H^+H^-$, and $WW'/ZZ'$ channels (if kinematically allowed). In what follows, we have defined $\xi_{HVV} = \frac{g_{HVV}}{g^{SM}_{hVV}}$ ($V = W^{\pm}, Z$), $\xi_{HVV^{'}} = \frac{g_{HVV^{'}}}{g^{SM}_{hVV}}$ ($V = W^{\pm}, Z$ and $V^{'} = W^{'\pm}, Z^{'}$, $\lambda_{HSS}$ is the Higgs self coupling ($S = A, h$), and finally $\xi_{Hf\bar{f}} = \frac{g_{Hf\bar{f}}}{g^{SM}_{hf\bar{f}}}$ and $N_{c} = 3\, (1)$ for the quarks (leptons). 
\begin{subequations}
\begin{align}
&\Gamma(H \rightarrow V V) = \frac{m^{3}_{H}}{16\pi v^{2}}\sqrt{1 - \frac{4m^{2}_{V}}{m^{2}_{H}}}\bigg(1 - \frac{4m^{2}_{V}}{m^{2}_{H}} + \frac{3}{4}\bigg(\frac{4m^{2}_{V}}{m^{2}_{H}}\bigg)^{2}\bigg)\xi^{2}_{HVV} \\
& \Gamma(H \rightarrow V V^{'}) = \frac{m^{2}_{V}m^{2}_{V^{'}}}{4\pi v^{2}m_{H}}\xi^{2}_{HVV^{'}}\bigg(2 + \frac{(m^{2}_{H} - m^{2}_{V^{'}} - m^{2}_{V})}{4m^{2}_{V}m^{2}_{V^{'}}}\bigg)\sqrt{1 - 2\bigg(\frac{m^{2}_{V} + m^{2}_{V^{'}}}{m^{2}_{H}}\bigg) + \bigg(\frac{m^{2}_{V^{'}} - m^{2}_{V}}{m^{2}_{H}}\bigg)^{2}}\\
& \Gamma(H \rightarrow SS) = \frac{\lambda^{2}_{HSS}}{32\pi m_{H}}\sqrt{1 - \frac{4m^{2}_{S}}{m^{2}_{H}}} \\
 &\Gamma(H \rightarrow H^{+}H^{-}) = \frac{\lambda^{2}_{HH^{+}H^{-}}}{32\pi m_{H}}\sqrt{1 - \frac{4m^{2}_{H^{\pm}}}{m^{2}_{H}}} \\
 &\Gamma(H \rightarrow f\bar{f}) = \frac{N_{c}m^{2}_{f}m_{H}}{8\pi v^{2}}\bigg(1 - \frac{4m^{2}_{f}}{m^{2}_{H}}\bigg)^{\frac{3}{2}}.
\end{align}
\end{subequations}
In Fig.~\ref{fig:HBR}, we plot the branching ratio of the $H$ into various channels for two different choices of $\sin\alpha$ and $\sin\beta$. While the qualitative picture seems largely consistent between the two plots, it is seen that the relative weights of the $Htt$ and the $WW/ZZ$ is quite different for the entire range of heavy higgs mass. This is due to the coupling $g_{HWW}$ being numerically smaller for the second choice, and $g_{Ht\bar{t}}\propto \cos\alpha$ being larger. The combined effect renders BR$(H\to t\bar{t})$ maximal thus also making it difficult experimentally to discover the heavy higgs for smaller values of the mixing angle $\sin\alpha$. However, there are other novel channels which in spite of lower decay branching ratios would nevertheless be of potentially great interest at the LHC. In terms of the collider signatures, we can deduce the following:

\begin{enumerate}
\item For a relatively light higgs ($m_H< 400$ GeV), in addition to the $tt$ channel (after it opens up kinematically), the most promising channels are the $WW$ and $ZZ$ corresponding to either multi-leptonic or multi-jet final states. The channel $H\to ZZ\to 4\ell$ with four hard leptons seems promising\footnote{The ATLAS search limit imposed in Fig.~\ref{fig:Hlimit} corresponds to the $\ell\ell j j$ channel.} as it aids in full reconstruction of the parent particle without any associated hadronic background, but comes with the price of tagging efficiency of the four leptons and the associated low $Z\to \ell\ell$ branching ratios. Also, in spite of a smaller branching ratio, the channel $H\to hh$ might also be viable - the presence of heavily boosted $b$ collinear quarks in the final state could invite employing fat jet techniques for reconstruction.
\item For a heavier higgs ($m_H> 400$ GeV), we see that in addition to the channels above (except the $hh$), there is the interesting possibility of it decaying into the charged higgs and a $W$, opening up interesting cascade decays. A particularly novel collider signature in this model is the decay chain $H\to H^+W^-$ with $H^+\to W'Z$ assuming the mass hierarchy $m_H>m_{H^+}>m_{W'}$. We discuss this and the $H^+\to WZ$ channel in more detail following Fig.~\ref{fig:BRlightcharged}.
\end{enumerate}
 
\subsection{The Charged Higgs $H^{\pm}$}

\begin{center}
 \begin{figure}[h!]
\includegraphics[scale = 0.8]{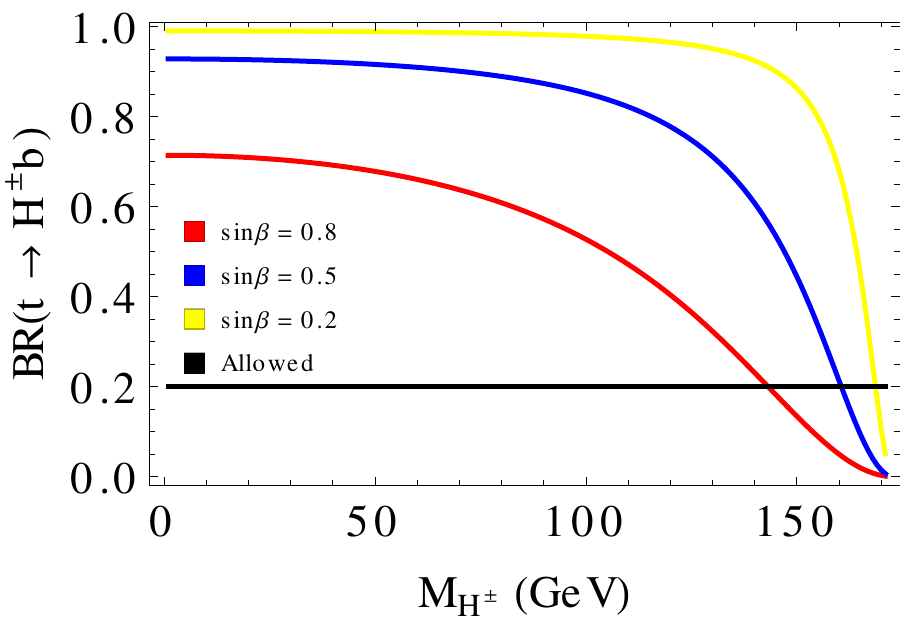}
\includegraphics[scale = 0.8]{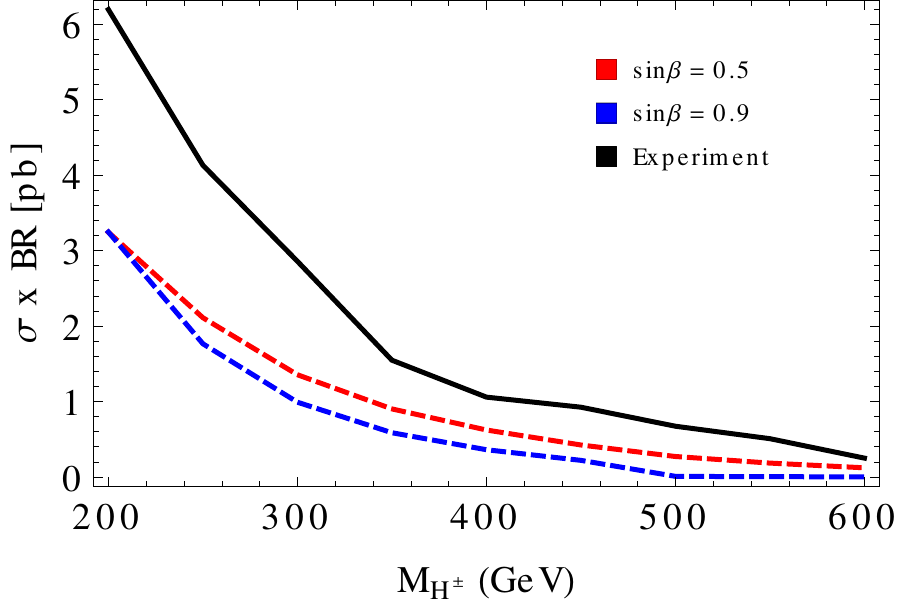}
\caption{(Left): The black dashed line shows the maximally allowed branching ratio for $t \rightarrow H^{\pm}b$ from the Tevatron experiments. It is seen that increasing the value of $\sin\beta$ can relax the lower mass bound for the charged Higgs, albeit only slightly. (Right): The LHC experimental exclusion limit in the channel $H^{\pm}\to t\bar{b}$ - we have translated the limits to the present model for two different values of $\sin\beta$.}
\label{fig:Hclimit}
\end{figure}
\end{center}

The charged higgs has been extensively looked for in the $tb$ and other channels for $m_{H^\pm}>m_t$ and in the $\tau\nu$ and $m_{H^\pm}<m_t$ whence it can be produced as a decay product of the top quark. The Tevatron D$\cancel{0}$ collaboration, has searched for the charged Higgs as a decay product of the top-quark, with the $H^{\pm}$ further decaying to either $c\bar{s}$ or $\tau^{\pm}\nu_{\tau}$ \cite{Abazov:2009aa}. The non-observation of the $H^{\pm}$ puts an upper bound on the branching ratio, BR$(t \rightarrow H^{\pm}t) \leq$ 0.12 - 0.26, depending upon the charged Higgs mass. In our model, the decay width of the top quark to charged Higgs and the bottom quark takes the form
\be  
\Gamma(t \rightarrow H^{\pm}  b) = \bigg(\frac{m_{t}}{16\pi v^{2}}\bigg)\big[m^{2}_{t}R^{2} + \mathcal{O}(m^{2}_{b})\big]\bigg(1 - \frac{m^{2}_{H^{\pm}}}{m^{2}_{t}}\bigg)^{2}, \nonumber
\ee
where
\be
R = \bigg(\frac{\lambda_{t}}{\lambda^{\textrm{SM}}_{t}}\bigg)\cos\beta = \cot\beta\bigg(\frac{2(1 - \frac{\epsilon^{2}_{L}}{2})M^{2}_{D} - 2m^{2}_{t}}{M^{2}_{D} - m^{2}_{t}}\bigg).
\label{eqn:R}
\ee
Note that for a sufficiently large Dirac mass, the quantity $R$ is insensitive to all the model parameters except $\sin\beta$ - in Fig.~\ref{fig:Hclimit} on the left, we display the branching ratio BR$(t \rightarrow H^{\pm}t)$ for various values of $\sin\beta$. Overlaid on the plot is a line corresponding to the BR value of 0.2\footnote{The precise bound of the BR depends on the $H^\pm$ mass - however, here we simply show the upper limit to extract the main qualitative features.} - it is seen that a charged Higgs with mass $m_{H^\pm}\leq 140$ GeV is mostly disallowed for the entire range of $\sin\beta$. Typically one would need very low values of $R$, or equivalently large $\sin\beta$, for a light charged Higgs to escape the Tevatron bounds.

The LHC collaborations have also looked for a heavy charged higgs for a wide range of masses in the $tb$ channel \cite{Aad:2015typ}, and have imposed upper limits on the $\sigma\times$BR value - on the right in Fig.~\ref{fig:Hclimit}, we display this limit (black curve) and also show the corresponding limit in our model for two values of $\sin\beta$. Given that a heavy charged higgs in our model has multiple decay channels, folding in the correct branching ratio to $tb$ gives a number that is well within the experimental bounds for the entire range of masses.

\begin{center} 
 \begin{figure}[h!]
\includegraphics[scale = 0.8]{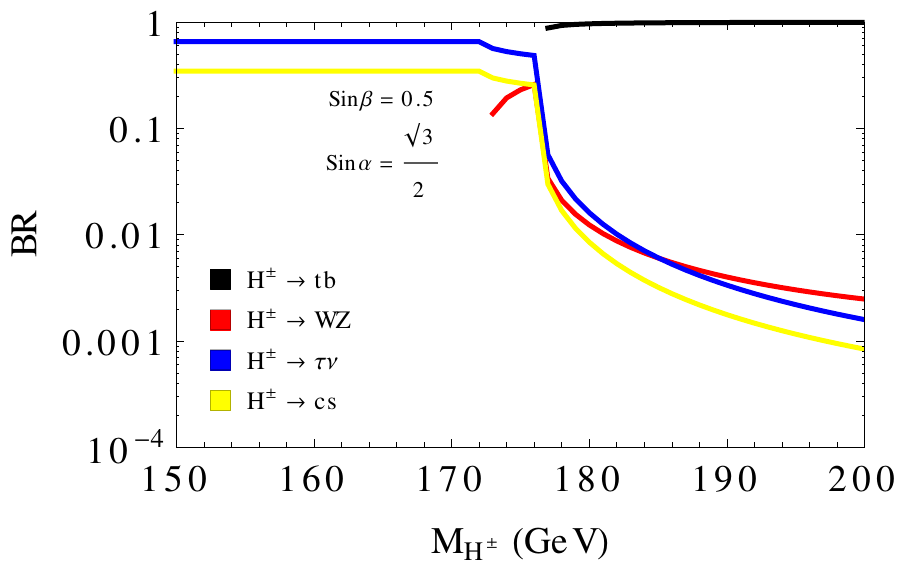}\hspace{0.1in}
\caption{In the left digram we have presented the branching ratio for the different channels which are kinematically allowed for the light charged Higgs. In the figure at right, The thick black curve showed the region excluded for the from the recent experimental searches conducted by both ATLAS and CMS.  }
\label{fig:BRlightcharged}
\end{figure}
\end{center}

To understand the best discovery modes for a charged higgs in this model, we begin with the light case ($m_{H^\pm}<$ 200 GeV) - in Fig.~\ref{fig:BRlightcharged}, we present the branching fractions into the various available channels. Before the $tb$ opens up, the decay is dominated by the $\tau\nu$ and the $cs$ channels. One interesting feature of this model is the presence of the $H^{\pm}WZ$ vertex, and thus the decay $H^{\pm}\to WZ$, though highly suppressed after the $tb$ becomes kinematically available, is still non-zero. There is a small interesting region around $m_{H^\pm}\approx$ 170 - 180 GeV where this decay is quite appreciable (with a 10-20\% BR). While the kinematic range in which this decay is appreciable is indeed very narrow, it can still serve as a useful discriminant from models in which the $SU(2)_L\times U(1)_Y$ breaking is carried out only by Higgs fields in the doublet representation (as in the 2HDM). In all such models, regardless of the number of higgs doublets, the $ZW^\pm H^\mp$ vertex is absent at tree level - see Ref.~\cite{Johansen:1982qm} for details.

The situation becomes quite interesting for more massive charged Higgses: in addition to the usual channels one probes, there are interesting new final states allowed in this model owing to the extended electroweak gauge sector. We begin by first listing out the partial decay widths in all the channels like in the previous heavy Higgs case.
\begin{subequations}
\begin{align}
&\Gamma(H^{\pm} \rightarrow V V^{'}) = \frac{m^{2}_{V}m^{2}_{V^{'}}}{4\pi v^{2}m_{H^{\pm}}}\xi^{2}_{H^{\pm}VV^{'}}\bigg(2 + \frac{(m^{2}_{H^{\pm}} - m^{2}_{V} - m^{2}_{V^{'}})}{4m^{2}_{V}m^{2}_{V^{'}}}\bigg)\sqrt{1 - 2\bigg(\frac{m^{2}_{V} + m^{2}_{V^{'}}}{m^{2}_{H^{\pm}}}\bigg) + \bigg(\frac{m^{2}_{V^{'}} - m^{2}_{V}}{m^{2}_{H^{\pm}}}\bigg)^{2}} \\
&\Gamma(H^{\pm} \rightarrow V S) = \frac{m^{3}_{H^{\pm}}}{8\pi v^{2}}\xi^{2}_{H^{\pm}VS}\bigg[1 - \frac{(m_{S} + m_{V})^{2}}{m^{2}_{H^{\pm}}}\bigg]^{\frac{3}{2}}\bigg[1 - \frac{(m_{V} - m_{S})^{2}}{m^{2}_{H^{\pm}}}\bigg]^{\frac{3}{2}}\\
&\Gamma(H^{\pm} \rightarrow f f^{'}) = \frac{N_{C}\lambda^{\frac{1}{2}}}{8\pi v^{2}m^{3}_{H^{\pm}}}\big[(m^{2}_{H^{\pm}} - m^{2}_{f}  - m^{2}_{f^{'}})(m^{2}_{f} + m^{2}_{f^{'}})\bigg(\frac{\lambda_{f}}{\lambda^{SM}_{f}}\bigg)^{2} - 4m^{2}_{f}m^{2}_{f^{'}} \big],
\end{align}
\end{subequations}
where the various  $\xi_{H^{\pm}VV'}$ are given by
\[ \xi_{H^{\pm}W^{'\mp}Z} = \frac{\sin\beta}{2}\bigg(1 + \frac{x^{2}}{4}\bigg);\,\, \xi_{H^{\pm}W^{\mp}Z} = \frac{x^{2}\cos\beta\sin\beta}{16\sin^{2}\theta_{w}\cos^{2}\theta_{w}};\,\,\xi_{H^{\pm}W^{\mp}Z^{'}} = \frac{\sin\beta\cos\theta_{w}}{2}\bigg(1 + \frac{x^{2}}{8}\bigg), \]
the $\xi_{H^{\pm}VS}$ by
\begin{align}
&\xi_{H^{\pm}W^{'\mp}A} = \frac{1}{x}\bigg(\sin^{2}\beta - \frac{x^{2}}{2}\bigg);\,\, \xi_{H^{\pm}W^{'\mp}h} = \frac{1}{\sqrt{2}x}\big[\sin\alpha\sin\beta - \frac{x^{2}}{32}(4\sin\alpha\cos\beta + \sqrt{2}\sin\alpha\sin\beta) \big];  \nonumber \\
& \xi_{H^{\pm}W^{\mp}A} = 1 + \frac{x^{2}}{32}(5 + 3\cos2\beta);\,\, \xi_{H^{\pm}W^{\mp}h} = \frac{1}{4}\big[(4\sin\alpha\cos\beta + \sqrt{2}\sin\alpha\sin\beta) + \frac{x^{2}}{32}(8\sin\alpha\cos\beta - \sqrt{2}\sin\alpha\sin\beta) \big],  \nonumber
\end{align}
and the Kallen function $ \lambda(m_{H^{\pm}}, m_{f}, m_{f^{'}}) = \big[m^{2}_{H^{\pm}} - (m_{f} + m_{f^{'}})^{2}\big]\big[m^{2}_{H^{\pm}} - (m_{f} - m_{f^{'}})^{2}\big]$.

\begin{center}
\begin{figure}[h!]
\includegraphics[scale=0.75]{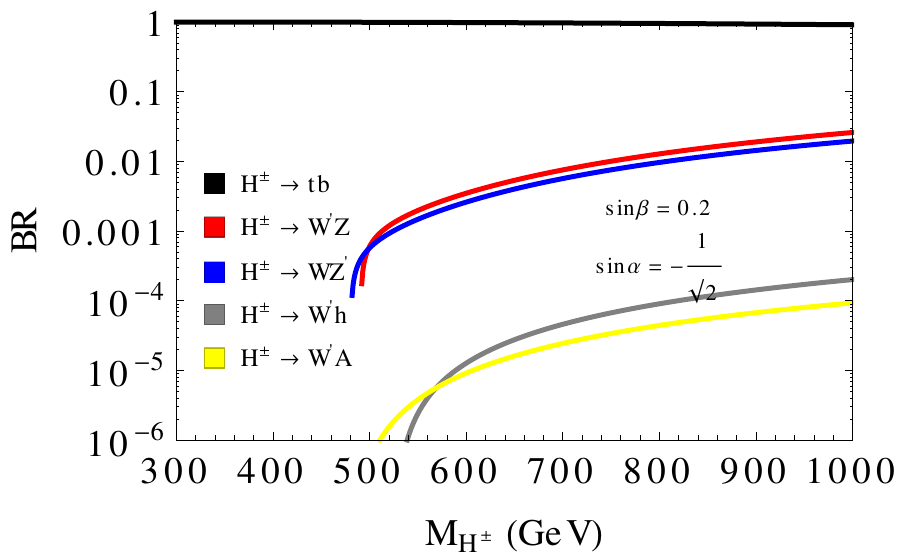}
\includegraphics[scale=0.75]{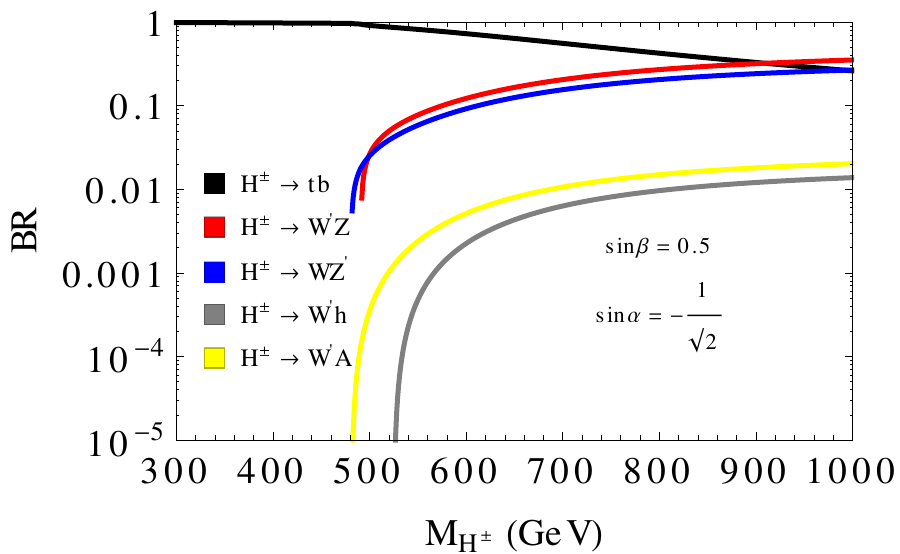}
\includegraphics[scale=0.75]{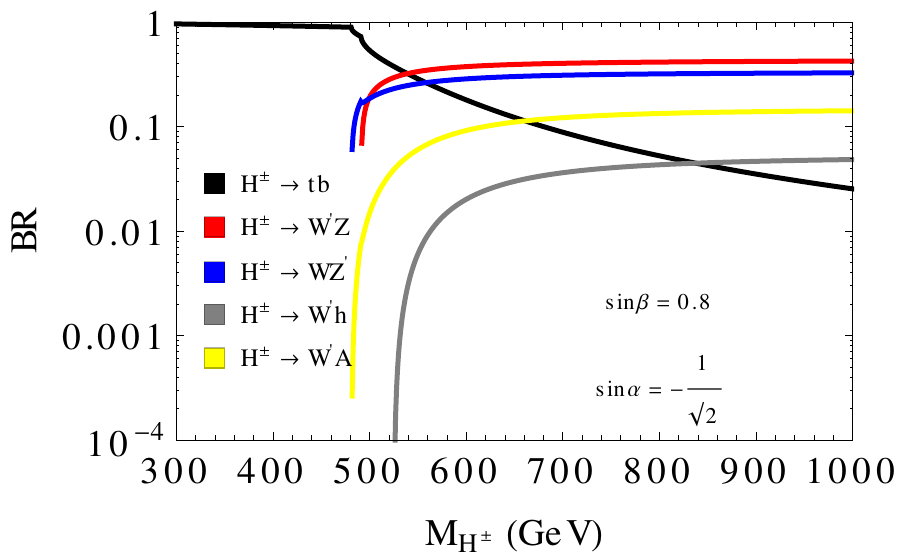}
\caption{In the above figure, we have presented the branching ratios for different dominating channels. From left to right we have shown the changing value of the branching ratio for the increasing value of $\sin\beta$ while keeping the $\sin\alpha$ value fix at $-0.7$.}
\label{fig:HpmBR}
\end{figure}
\end{center}

In Fig.~\ref{fig:HpmBR}, we display the branching ratio of a heavy charged Higgs ($m_{H^\pm}>300$ GeV) into various channels for three different values of $\sin\beta$ with $\sin\alpha$ fixed at $-0.7$.\footnote{The reason for this choice is that $\sin\alpha=-0.7$ allows for a wide range of $\sin\beta$ values - see Fig.~\ref{fig:h125}.} and for $M_{W'}=400$ GeV.
Certain qualitative features of these plots are worth noting.
\begin{itemize}
\item For small values of $\sin\beta$, the $H^+tb$ coupling becomes large (see Eqn.~\ref{eqn:R}), and hence this decay mode dominates over the entire range of masses.
\item For moderate values of $\sin\beta$, there are other competing modes that open up beyond $M_{H^\pm}\approx 750$ GeV. It is interesting to note that the presence of a light $W'$ (with mass 400 GeV as chosen) facilitates the discovery of the charged Higgs via the channel $H^+\to W'^+ Z$. Since the $W'$ in our model has no couplings to SM fermions at tree level, we are left with the decay chain $H^{\pm}\to W'^{\pm} Z\to W^{\pm}ZZ$ with a potentially interesting final state topology with multiple hard leptons and/or jets.
\item The $H^{\pm}W^{\mp}Z'$ coupling has an additional $\cos\theta_w$ factor compared to $\xi_{H^{\pm}W^{'\mp}Z}$ and thus is subdominant in the entire parameter space. However, the final decay chain for the process $H^{\pm}\to W^{\pm} Z'\to W^{\pm}ZZ$ could also be potentially probed at the LHC. For both these processes, there could also be interesting cascade decays possible with the heavy vector boson decaying to a higgs and a lighter vector boson.
\end{itemize}
\section{Conclusions}
\label{sec:conculsions}
In this paper, we presented a gauge extension of the SM inspired by the ``Top Triangle Moose" construction in Ref.~\cite{Chivukula:2009ck} and modified it minimally by replacing one of the two non-linear sigma models with a Higgs field. The pattern of symmetry breaking pattern aided by both Higgs and non-linear sigma model fields results in a scalar spectrum that is identical to the well-known 2HDM class in spite of markedly different gauge and scalar sectors. The particle content includes heavy vector, scalar, and fermionic resonances in addition to the SM particles. We have chosen to construct the model so that it mimics the Type I 2HDM in terms of the pattern of the neutral higgs couplings to the SM fermions. 

The heavy and light neutral higgses in the model, while coupling differently to the SM $W,Z$ exhibit a different alignment limit compared to the 2HDM. We have imposed the constraints coming from the 125 GeV higgs discovery and find that there is a fair amount of parameter space that can potentially be directly probed at the LHC. The model retains the  ideal fermion delocalization aspect of ``higgsless" models that came to fore prior to the higgs discovery effectively rendering the heavy gauge bosons fermiophobic. Taken together with the fact that the heavy neutral and charged higgses can decay into the $W'$ and $Z'$, we find that this model admits many striking collider signatures that involve multiple SM gauge bosons in the final state. The presence of the additional $SU(2)$ group affects the higgs phenomenology in interesting ways, a particular example being the $WZ$ and $W'Z$ decay modes of the charged Higgs that are absent in 2HDM and other higgs doublet models. 

This work (re)emphasizes that it is always possible to construct models which have distinct UV-completion but apparently look indistinguishable at low energy regime. Favoring or rejecting a particular class of models from others merely considering the discovery of a new particle in the future collider experiments might not be a viable tactic going forward as many of these do have parameter spaces wherein the cross-section numbers look very similar. For a successful ``inverse collider program", one should simultaneously test a theory from multiple experimental fronts. This work reinforces that the future paradigm of the collider experiments is one where both the energy and precision frontiers coexist on an equal standpoint. In that spirit, we will undertake the full collider study of this model in an upcoming work.                                          
\begin{acknowledgments}
We thank Sekhar Chivukula and Elizabeth Simmons for a careful reading of the manuscript and providing useful suggestions.
\end{acknowledgments}
\appendix
\section{Gauge and Fermion Couplings}
\label{app: gauge}
We will start by calculating all the self-couplings to $\mathcal{O}(x^{2})$. We start with an example each of triple gauge vertex and a four point coupling:
\begin{align*}
G_{WWZ} & = g\big(v^{W}_{0}\big)^{2}v^{Z}_{0} + g_{1}\big(v^{W}_{1}\big)^{2}v^{Z}_{1} \\
                  & = - \frac{e\cos\theta_{w}}{\sin\theta_{w}}\bigg(1 + \frac{x^{2}}{8\cos^{2}\theta_{w}}\bigg),
\end{align*} 
\begin{align*}
G_{WWZ\gamma} & = g^{2}\big(v^{W}_{0}\big)^{2}v^{Z}_{0}\big(\frac{e}{g}\big) + g_{1}^{2}\big(v^{W}_{1}\big)^{2}v^{Z}_{1}\big(\frac{e}{g_{1}}\big)  \\
      & = e\big[ g\big(v^{W}_{0}\big)^{2}v^{Z}_{0} + g_{1}\big(v^{W}_{1}\big)^{2}v^{Z}_{1} \big] \\ 
      & = eG_{WWZ}.
\end{align*} 
Here the $v^{Z}_{i}$ etc.\ are the amount of overlap of the particular gauge boson in site $i$ - see Ref.~\cite{Chivukula:2009ck,Chivukula:2011ag} . In a similar fashion, we can calculate all three point and four point couplings - we list the results in Tables~\ref{tab:three-point} and \ref{tab:four-point}. 
\begin{table}[h!]
\begin{center}
{
\begin{tabular}{|   c   ||   c   || c  |}
\hline
Vertex & Computed as & Strength \\
\hline \hline
$G_{WW\gamma}$ & $g(v^{W}_{0})^{2}v^{\gamma}_{0} + g_{1}(v^{W}_{1})^{2}v^{\gamma}_{1}$ & $e$ \\
\hline
$G_{W^{'}W^{'}\gamma}$ & $g(v^{W^{'}}_{0})^{2}v^{\gamma}_{0} + g_{1}(v^{W^{'}}_{1})^{2}v^{\gamma}_{1}$ & $e$ \\
\hline
$G_{WW^{'}\gamma}$ & $gv^{W}_{0}v^{W^{'}}_{0}v^{\gamma}_{0} + g_{1}v^{W}_{1}v^{W^{'}}_{1}v^{\gamma}_{1}$ & 0 \\
\hline
$G_{WWZ}$ & $g(v^{W}_{0})^{2}v^{Z}_{0} + g_{1}(v^{W}_{1})^{2}v^{Z}_{1}$ & $- \frac{e\cos\theta_{w}}{\sin\theta_{w}}\bigg[1 + \frac{x^{2}}{8\cos^{2}\theta_{w}}\bigg]$ \\
\hline
$G_{WWZ^{'}}$ & $g(v^{W}_{0})^{2}v^{Z^{'}}_{0} + g_{1}(v^{W}_{1})^{2}v^{Z^{'}}_{1}$ & $\frac{ex}{4\sin\theta_{w}}$ \\
\hline
$G_{W^{'}W^{'}Z^{'}}$ & $g(v^{W^{'}}_{0})^{2}v^{Z^{'}}_{0} + g_{1}(v^{W^{'}}_{1})^{2}v^{Z^{'}}_{1}$ &$- \frac{e}{x\sin\theta_{w}}\bigg[1 - \frac{x^{2}}{8}\big(3 + \sec^{2}\theta_{w}\big)\bigg]$ \\
\hline
$G_{W^{'}W^{'}Z}$ & $g(v^{W^{'}}_{0})^{2}v^{Z}_{0} + g_{1}(v^{W^{'}}_{1})^{2}v^{Z}_{1}$ & $- \frac{e}{2\sin\theta_{w}\cos\theta_{w}}\bigg[1 + \frac{x^{2}}{8}\sec^{2}\theta_{w} - 2\cos^{2}\theta_{w}\bigg]$ \\
\hline
$G_{WW^{'}Z}$ & $gv^{W}_{0}v^{W^{'}}_{0}v^{Z}_{0} + g_{1}v^{W}_{1}v^{W^{'}}_{1}v^{Z}_{1}$ & $\frac{ex}{4\sin\theta_{w}\cos\theta_{w}}$ \\
\hline
$G_{WW^{'}Z^{'}}$ & $gv^{W}_{0}v^{W^{'}}_{0}v^{Z^{'}}_{0} + g_{1}v^{W}_{1}v^{W^{'}}_{1}v^{Z^{'}}_{1}$ & $- \frac{e}{2\sin\theta_{w}}\bigg[1 + \frac{x^{2}}{2}\big(1 + \frac{\sec^{2}\theta_{w}}{4}\big)\bigg]$ \\
\hline
\end{tabular}
}
\caption{Triple gauge boson couplings in the model - these have been evaluated perturbatively in $x$ to $\mathcal{O}(x^2)$.}
\label{tab:three-point}
\end{center}
\end{table}

\begin{table}[h!]
\begin{center}
{
\begin{tabular}{|   c   ||   c   |}
\hline
Vertex & Strength \\
\hline \hline
$G_{WW\gamma\gamma}$ & $e^{2}$ \\
\hline
$G_{W^{'}W^{'}\gamma\gamma}$ & $e^{2}$ \\
\hline
$G_{WW^{'}\gamma\gamma}$ & $ 0 $ \\
\hline
$G_{WWZ\gamma}$ & $e\,G_{WWZ}$ \\
\hline
$G_{WWZ^{'}\gamma}$ & $e\,G_{WWZ^{'}}$ \\
\hline
$G_{WW^{'}Z\gamma}$ & $e\,G_{WW^{'}Z}$ \\
\hline
$G_{WW^{'}Z^{'}\gamma}$ & $e\,G_{WW^{'}Z^{'}}$ \\
\hline
$G_{W^{'}W^{'}Z\gamma}$ & $e\,G_{W^{'}W^{'}Z}$ \\
\hline
$G_{W^{'}W^{'}Z^{'}\gamma}$ & $e\,G_{W^{'}W^{'}Z^{'}}$ \\
\hline
\end{tabular}
}
\caption{Four point gauge couplings in the model - these have been evaluated perturbatively in $x$ to $\mathcal{O}(x^2)$.}
\label{tab:four-point}
\end{center}
\end{table}
The gauge-fermion couplings can be calculated similarly and we list the charged current and neutral current interactions in Tables~\ref{tab:charged} and \ref{tab:neutral}.
\begin{table}[h!]
\begin{center}
\renewcommand{\arraystretch}{2}
\resizebox{9cm}{!}
{
\begin{tabular}{|   c   ||   c   ||   c   |}
\hline 
Coupling & Computed As & Strength \\
\hline \hline
$g^{Wtb}_{L}$ & $gv^{0}_{W}t^{0}_{L}b^{0}_{L} + g_{1}v^{1}_{W}t^{1}_{L}b^{1}_{L}$ & $\frac{e}{\sin\theta_{w}}$ \\ 
\hline
$g^{WTb}_{L} = g^{WtB}$ & $gv^{0}_{W}T^{0}_{L}b^{0}_{L} + g_{1}v^{1}_{W}T^{1}_{L}b^{1}_{L}$ & $- \frac{ex}{2\sqrt{2}\sin\theta_{w}}$ \\
\hline
$g^{WTB}_{L}$ & $gv^{0}_{W}T^{0}_{L}B^{0}_{L} + g_{1}v^{1}_{W}T^{1}_{L}B^{1}_{L}$ & $\frac{e}{2\sin\theta_{w}}\big(1 + \frac{3}{8}x^{2}\big)$ \\
\hline
$g^{Wtb}_{R}$ & $g_{1}v^{1}_{W}t^{1}_{R}b^{1}_{R}$ & 0 \\
\hline
$g^{WTb}_{R} = g^{WtB}_{R}$ & $g_{1}v^{1}_{W}T^{1}_{R}b^{1}_{R}$ & 0 \\
\hline 
$g^{WTB}_{R}$ & $g_{1}v^{1}_{W}T^{1}_{R}B^{1}_{R}$ & $\frac{e}{2\sin\theta_{w}}\big(1 - \frac{x^{2}}{8}\big)$ \\
\hline
$g^{W^{'}tb}_{L}$ & $gv^{0}_{W^{'}}t^{0}_{L}b^{0}_{L} + g_{1}v^{1}_{W^{'}}t^{1}_{L}b^{1}_{L}$ & 0 \\
\hline
$g^{W^{'}Tb}_{L} = g^{W^{'}tB}_{L}$ & $gv^{0}_{W^{'}}T^{0}_{L}b^{0}_{L} + g_{1}v^{1}_{W^{'}}T^{1}_{L}b^{1}_{L}$ & $\frac{e}{\sqrt{2}\sin\theta_{w}}$ \\
\hline
$g^{W^{'}TB}_{L}$ & $gv^{0}_{W^{'}}T^{0}_{L}B^{0}_{L} + g_{1}v^{1}_{W^{'}}T^{1}_{L}B^{1}_{L}$ & $\frac{e}{x\sin\theta_{w}}\big(1 - \frac{3}{4}x^{2}\big)$ \\
\hline
$g^{W^{'}tb}_{R}$ & $g_{1}v^{0}_{W}t^{1}_{R}b^{1}_{R}$ & 0 \\
\hline
$g^{W^{'}Tb} = g^{W^{'}tB}$ & $g_{1}v^{0}_{W}T^{1}_{R}b^{1}_{R}$ & 0 \\
\hline
$g^{W^{'}TB}_{R}$ & $g_{1}v^{1}_{W^{'}}T^{1}_{R}B^{1}_{R}$ & $\frac{e}{x\sin\theta_{w}}\big(1 - \frac{x^{2}}{4}\big)$ \\
\hline
\end{tabular}
}
\caption{Charged current interactions in the model calculated to $\mathcal{O}(x^2)$.}
\label{tab:charged}
\end{center}
\end{table}
\begin{table}
\begin{center}
\renewcommand{\arraystretch}{3}
\resizebox{12cm}{!}
{
\begin{tabular}{|   c   ||   c   ||   c   |}
\hline 
Coupling & Computed As & Strength \\
\hline \hline
$g^{Ztt}_{L}$ & $(gv^{0}_{Z}(t^{0}_{L})^{2} + g_{1}v^{1}_{Z}(t^{1}_{L})^{2})T_{3} + g_{2}v^{2}_{Z}((t^{0}_{L})^{2} + (t^{1}_{L})^{2}))(Q - T_{3})$ & $- \frac{e}{\sin\theta_{w}\cos\theta_{w}}(T_{3} - \sin^{2}\theta_{w}Q)$ \\
\hline
$g^{ZtT}_{L}$ & $(gv^{0}_{Z}t^{0}_{L}T^{0}_{L} + g_{1}v^{1}_{Z}t^{1}_{L}T^{1}_{L}) + g_{2}v^{2}_{Z}(t^{0}_{L}T^{0}_{L} + t^{1}_{L}T^{1}_{L})(Q - T_{3})$ & $\frac{exT_{3}}{2\sqrt{2}\sin\theta_{w}\cos\theta_{w}}$ \\
\hline
$g^{ZTT}_{L}$ & $(gv^{0}_{Z}(T^{0}_{L})^{2} + g_{1}v^{1}_{Z}(T^{1}_{L})^{2}) + g_{2}v^{2}_{Z}((T^{0}_{L})^{2} + (T^{1}_{L})^{2})(Q - T_{3})$ & $- \frac{e}{\sin\theta_{w}\cos\theta_{w}}\bigg(\frac{1}{2}\big[1 + \frac{x^{2}}{8}(4 - \sec^{2}\theta_{w})\big]T_{3} - Q\sin^{2}\theta_{w}\bigg)$ \\
\hline
$g^{Ztt}_{R}$ & $g_{1}v^{1}_{Z}(t^{1}_{R})^{2}T_{3} + g_{2}v^{2}_{Z}((t^{1}_{R})^{2} + (t^{2}_{R})^{2})(Q - T_{3})$ & $e\tan\theta_{w}(Q - T_{3})$ \\
\hline
$g^{ZtT}_{R}$ & $g_{1}v^{1}_{Z}(t^{1}_{R}T^{1}_{R})T_{3} + g_{2}v^{2}_{Z}((t^{1}_{R}T^{1}_{R}) + (t^{2}_{R}T^{2}_{R}))(Q - T_{3})$ & $0$ \\
\hline
$g^{ZTT}_{R}$ & $g_{1}v^{1}_{Z}(T^{1}_{R})^{2}T_{3} + g_{2}v^{2}_{Z}((T^{1}_{R})^{2} + (T^{2}_{R})^{2})(Q - T_{3})$ & $- \frac{e}{\sin\theta_{w}\cos\theta_{w}}\bigg(\frac{1}{2}\big[(1 - \frac{x^{2}}{8}\sec^{2}\theta_{w})T_{3}\big] - Q\sin^{2}\theta_{w}\bigg)$ \\
\hline
$g^{Z^{'}tt}_{L}$ & $(gv^{0}_{Z^{'}}(t^{0}_{L})^{2} + g_{1}v^{1}_{Z^{'}}(t^{1}_{L})^{2})T_{3} + g_{2}v^{2}_{Z^{'}}((t^{0}_{L})^{2} + (t^{1}_{L})^{2}))(Q - T_{3})$ & $\frac{ex\tan\theta_{w}}{2\cos\theta_{w}}(Q - T_{3})$ \\
\hline
$g^{Z^{'}tT}_{L}$ & $(gv^{0}_{Z^{'}}t^{0}_{L}T^{0}_{L} + g_{1}v^{1}_{Z^{'}}t^{1}_{L}T^{1}_{L}) + g_{2}v^{2}_{Z^{'}}(t^{0}_{L}T^{0}_{L} + t^{1}_{L}T^{1}_{L})(Q - T_{3})$ & $- \frac{e}{\sqrt{2}\sin\theta_{w}}\bigg(1 - \frac{x^{2}}{8}\tan^{2}\theta_{w}\bigg)T_{3}$ \\
\hline
$g^{Z^{'}TT}_{L}$ & $(gv^{0}_{Z^{'}}(T^{0}_{L})^{2} + g_{1}v^{1}_{Z^{'}}(T^{1}_{L})^{2}) + g_{2}v^{2}_{Z^{'}}((T^{0}_{L})^{2} + (T^{1}_{L})^{2})(Q - T_{3})$ & $- \frac{e}{x\sin\theta_{w}}\bigg[1 - \frac{x^{2}}{8}\big(4 - 3\tan^{2}\theta_{w}\big)\bigg]T_{3} + \frac{ex\tan\theta_{w}}{2\cos\theta_{w}}Q$ \\
\hline
$g^{Z^{'}tt}_{R}$ & $g_{1}v^{1}_{Z^{'}}(t^{1}_{R})^{2}T_{3} + g_{2}v^{2}_{Z^{'}}((t^{1}_{R})^{2} + (t^{2}_{R})^{2})(Q - T_{3})$ & $\frac{ex\tan\theta_{w}}{2\cos\theta_{w}}(Q - T_{3})$ \\
\hline
$g^{Z^{'}tT}_{R}$ & $g_{1}v^{1}_{Z^{'}}(t^{1}_{R}T^{1}_{R})T_{3} + g_{2}v^{2}_{Z^{'}}((t^{1}_{R}T^{1}_{R}) + (t^{2}_{R}T^{2}_{R}))(Q - T_{3})$ & $0$ \\
\hline
$g^{Z^{'}TT}_{R}$ & $g_{1}v^{1}_{Z}(T^{1}_{R})^{2}T_{3} + g_{2}v^{2}_{Z}((T^{1}_{R})^{2} + (T^{2}_{R})^{2})(Q - T_{3})$ & $- \frac{e}{x\sin\theta_{w}}\bigg[1 - \frac{3x^{2}}{8}\tan^{2}\theta_{w}\bigg]T_{3} + \frac{ex\tan\theta_{w}}{2\cos\theta_{w}}Q$ \\
\hline
\end{tabular}
}
\caption{Neutral current interactions in the model calculated to $\mathcal{O}(x^2)$.}
\label{tab:neutral}
\end{center}
\end{table}
\clearpage

\bibliography{References}

\end{document}